\documentstyle[eqsecnum,pre,preprint,aps,epsfig]{revtex}

\tightenlines

\newcommand{\bq}{\begin{equation}}
\newcommand{\eq}{\end{equation}}
\newcommand{\bqa}{\begin{eqnarray}}
\newcommand{\eqa}{\end{eqnarray}}

\def\prep#1#2#3{ Phys. Rep. ${\bf{#1}}$ (#2) #3}
\def\etal{{\it et.al.\/}}

\begin{document}

\draft
\preprint{PM/00-23, corrected version}

\title{Logarithmic SUSY electroweak effects on four-fermion
processes at TeV energies}

\author{M. Beccaria$^{a,b}$, F.M.
Renard$^c$ and C. Verzegnassi$^{d,e}$ \\
\vspace{0.4cm} 
}

\address{
$^a$Dipartimento di Fisica, Universit\`a di 
Lecce \\
Via Arnesano, 73100 Lecce, Italy.\\
\vspace{0.2cm}  
$^b$INFN, Sezione di Lecce\\
Via Arnesano, 73100 Lecce, Italy.\\
\vspace{0.2cm} 
$^c$ Physique
Math\'{e}matique et Th\'{e}orique, UMR 5825\\
Universit\'{e} Montpellier
II,  F-34095 Montpellier Cedex 5.\hspace{2.2cm}\\
\vspace{0.2cm} 
$^d$
Dipartimento di Fisica Teorica, Universit\`a di Trieste, \\
Strada Costiera
 14, Miramare (Trieste) \\
\vspace{0.2cm} 
$^e$ INFN, Sezione di Trieste\\
}

\maketitle

\begin{abstract}
We compute the MSSM one-loop contributions to the asymptotic energy 
behaviour  of fermion-antifermion pair production at future 
lepton-antilepton colliders. Besides the conventional logarithms of 
Renormalization Group
origin, extra SUSY linear logarithmic terms appear of "Sudakov-type". 
In the TeV range their overall effect on a variety of observables
can be quite relevant and  drastically different from that       
obtained in the SM case. 
\end{abstract}

\pacs{PACS numbers: 12.15.-y, 12.15.Lk, 14.65.Fy, 14.80.Ly}

\section{Introduction.} 

In recent papers \cite{log}, \cite{mt}, 
the effects of one-loop diagrams on 
fermion-antifermion pair
production at future lepton-antilepton colliders were computed in the SM
for both massless \cite{log} and massive (in practice, 
bottom production) \cite{mt} fermions.
As a result of that calculation it was found that, in the high energy
region, contributions arise that are both of linear and of quadratic
logarithmic kind in the c.m. energy, but are not of 
Renormalization Group (RG)
origin.
For this reason they were called \cite{CC}
"of Sudakov-type", \cite{Sudakov}, 
although the
theoretical mechanism that generates them is not, rigorously speaking,
of infrared origin, as exaustively discussed in following articles
\cite{CCC}.
In this paper, we shall retain the original "Sudakov-type" notation,
but one might call these terms e.g. "not of RG origin" to avoid
theoretical confusion.\\

As a by product of our computations, it was also stressed in \cite{mt}
that, in the
special case of bottom-antibottom production, extra terms appear that
are "of Sudakov-type" and also quadratic in the top mass, a situation
that reminds that met at the $Z$-peak in the calculation of the partial
$Z$ width into $b\bar b$. Neglecting these terms would produce a
serious theoretical mistake in the case of certain observables,
particularly the $b\bar b$ cross section, and in principle (for very
high lumonisity) also in the $b\bar b$ longitudinal polarization
asymmetry.\\

When the c.m. energy crosses the typical TeV limit, the relative
effects of the "Sudakov-type" logarithms begin to rise well beyond the
(tolerable) few percent threshold, making the validity of a one-loop
approximation not always obvious, depending on the chosen observable.
In particular, hadronic production seems to be in a critical shape, as
discussed in \cite{CCC}. These conclusions are quite
different from those that would be obtained if only the RG linear
asymptotic logarithms were retained. In that case, the smooth relative
effect would remain systematically under control at TeV energies, not
generating special theoretical diseases. On the contrary, in the
"Sudakov" case a subtle mechanism of opposite linear and quadratic
logarithms contributions often appears that makes the overall effect
less controlable. Thus, neglecting the non RG asymptotic effects in the
considered processes would certainly be a theoretical disaster.\\

The aim of this paper is that of investigating whether similar
conclusions can be drawn when one works in the framework of a
supersymmetric extension of the SM. In particular, although the same
analysis could be performed in a more general case, we shall fix our
attention here on the simplest minimal SUSY model (MSSM) \cite{MSSM}. 
We shall be
motivated in this search by (at least) two qualitative reasons. These
are a consequence of the results obtained in Ref.\cite{log}, 
showing that in
some cases the relative size of the effects becomes larger than the
expected experimental accuracy. If SUSY extra diagrams increased this
value, their rigorous inclusion at one-loop would be essential e.g. for
a test of the theory if SUSY partners were discovered. But even if
direct production were still lacking for some special "heavy" SUSY
particles (e.g. neutralinos), a large virtual effect in some observable
might be, in principle, detectable. In this spirit, we shall proceed in
this paper as follows. We shall assume that SUSY has been at least
partially detected, and that for all the masses of the model a "natural"
mechanism \cite{nat} exists that confines their values below the TeV
limit (in practice, they might roughly be of the same size as the top
mass). In this spirit, the c.m. energy region beyond one TeV can be
considered as "nearly" asymptotic. This means that we shall have in our
minds, more than the future 500 GeV Linear Collider (LC) 
\cite{LC} case, that
of the next CERN Compact Linear Collider (CLIC) \cite{CLIC}, 
supposed to be
working at energies between 3 and 5 TeV. With due care, though, we feel
that a number of our conclusions might well be extrapolated to the LC
situation, as illustrated in the original Ref.\cite{log}.\\

In Section II of this paper we shall review the various MSSM
diagrams that give rise to "Sudakov" logarithms and discuss the
analogies and the differences with respect to the SM. We
shall discuss separately the various contributions both in the massless
case and in that of $b\bar b$ production (the case of top production, 
that requires
a modification of the adopted theoretical 
scheme, will be   
treated separately in a forthcoming paper). For final bottom, 
we shall show 
that the
overall logarithmic genuine SUSY contributions that are also quadratic
in the top mass enhance the
corresponding SM ones.
Moreover, there appear terms that are quadratic in the bottom mass
and are multiplied by $\tan^2\beta$, which could also be sizeable
for very large values of $\tan^2\beta$.  
The obtained expressions of the various observables will be
shown in Section III, 
and the features of the MSSM relative effects will be
displayed in several Figures. It will appear that the MSSM logarithmic
effects are drastically different from those of the SM, and again quite
different from those obtained in the pure RG approximation. The
expectable validity of a logarithmic parametrization will be
discussed in the final Section IV, with special emphasis on the CLIC
energy region but also on the LC case. The possibility of a relatively
simple parametrization to be used in the TeV energy range will be
also qualitatively motivated. Finally,
a short Appendix will contain the detailed 
asymptotic logarithmic contributions from various diagrams to
the four gauge-invariant functions that in our 
approach generate all the observable quantities of the 
considered process at the electroweak one loop.

\section{MSSM diagrams generating asymptotic logarithms}

The theoretical analysis of this paper is based on the use 
of the so called "$Z$-peak-subtracted"
representation, which has been illustrated in several previous 
references \cite{Zsub} and was 
conveniently used to describe the process of electron-positron 
annihilation into a final fermion ($f$) antifermion
$\bar f$, that can be either a
lepton-antilepton or a "light" ($u,d,s,c,b$) quark-antiquark pair.
For what concerns the genuine
electroweak sector of the process, all the relevant information 
is provided by four gauge-invariant
functions of $q^2$ and $\theta$ (the squared c.m. energy and 
scattering angle) that are called 
$\tilde{\Delta}_{\alpha lf}$, $R_{lf}$, $V^{\gamma Z}_{lf}$, 
$V^{Z\gamma}_{lf}$ and describe one-loop transitions
of various Lorentz structure (photon-photon, $Z$-$Z$, photon-$Z$ 
and $Z$-photon respectively). These functions vanish
by construction at $q^2 =0$ 
($\tilde{\Delta}_\alpha$) and $q^2$= $M_Z^2$ 
(the other three quantities) respectively and
are ultraviolet finite. They enter the theoretical expression 
of the various cross sections and asymmetries
in a way that is summarized in the  Appendix B of Ref.\cite{log}, 
and we will not insist on their properties here.\\

At one loop, the previous four gauge-invariant functions 
receive contributions from diagrams of self-energy,
vertex and box type. Self-energy diagrams with a small 
addition of the "pinch" part of the $WW$ vertex
generate asymptotically logarithms of the c.m. energy in
agreement with the 
Renormalization Group (RG) treatment. 
Extra logarithms of "pseudo-Sudakov" type (we follow the original 
denomination of Degrassi and Sirlin \cite{DS}, 
whose description of four-fermion processes 
has been adopted in our work) arise in the SM from
two kinds of diagrams. Vertices with one or two internal $W's$ or 
one internal $Z$ generate both linear and quadratic logarithms;
boxes with either $W's$ or $Z's$ do the same.
For massless fermions, there are no other types of logarithms.
However, for final bottom-antibottom production, 
vertex diagrams produce extra linear logarithms that are
also quadratic in the top mass, and cannot be neglected. 
All these results can be found in \cite{log}, \cite{mt}; for
completeness we have also written the same type of terms quadratic in 
the bottom mass although they are numerically negligible.\\

When one moves to the MSSM, the situation becomes, 
at least for what concerns this special topics,
relatively simpler. In fact, one discovers immediately 
that box diagrams with internal SUSY partners
do not generate asymptotic
logarithms. This feature, that is quite different from the SM one, 
is due to the different spin structure of the fermion-fermion-scalar
couplings which arise in SUSY and replace the fermion-fermion-vector
couplings arising in SM. As a consequence, when the energy increases,
the SUSY box contribution vanish as an inverse power of $q^2$.
Thus only self-energies and vertices must be considered. 
Self-energies  will generate the 
RG logarithmic behaviour. Summing the various bubbles involving
SUSY partners ($\tilde{f}$, $\chi^{\pm}$, $\chi^0$), 
Higgses ($A^0$, $H^0$, $h^0$),  and Goldstones,
we obtain the self-energy contributions to the four
functions $\tilde{\Delta}_{\alpha lf}$, $R_{lf}$, $V^{\gamma Z}_{lf}$, 
$V^{Z\gamma}_{lf}$ given in the Appendix. Using the relations between
these contributions and the expressions giving the running of
$g_1$, $g_2$, $s^2_W$ established in Ref.\cite{log}, we have checked
that our result agrees with the running quoted in the literature
\cite{GUT} for both the SM and the MSSM cases.\\

For vertices, the analysis is, to our knowledge, new and, 
in our opinion, interesting. 
First, and again because of the absence of helicity conserving
fermion-fermion-vector couplings, in SUSY there is no helicity structure
analogue to the one brought by the SM ($WWf$) triangle and then 
no quadratic logarithmic contribution.
However there appears linear logarithmic contributions called
of "Sudakov-type" because they are not universal and do not contribute
to RG. For massless fermions, they are generated by the 
diagrams that involve chargino(s) or neutralino(s) 
together with sfermions exchanges as shown in Fig.1; 
the related effects on the four functions are given in the
Appendix. They can be compared with the corresponding SM effects 
computed in Ref.\cite{log}, Section 2.3.\\

A special discussion is due to the case of final $b\bar b$ 
production. Here to 
the previous SUSY diagrams one must
add the contributions from the MSSM Higgses, exactly like 
in the SM case. So we shall have both contributions
of chargino/neutralino-sfermion origin, see Fig.1 
(denoted by a symbol $\chi$), and of Higgs 
origin, see Fig.2 (denoted by a symbol $H$). 
Note that, being interested in the 
additional contribution brought by SUSY, to be later on added to the
SM contribution in order to get the full MSSM one,
in the Higgs part ($H$), we write
the total MSSM Higgs contribution minus the SM Higgs contribution.\\

For the purposes of the following discussion it is convenient 
to write the effects of the previous diagrams,
rather than on the gauge-invariant subtracted functions, 
on the photon and $Z$ vertices $\Gamma_\mu ^{\gamma}$,
$\Gamma_\mu ^Z$, defined in a conventional way \cite{log},\cite{DS}. 
One easily finds first
the $\chi$ contribution and secondly the ($H$)
contribution:

\bqa
\Gamma^{\gamma}_{\mu}(\chi)&\to&
-({e\alpha\over48\pi M^2_Ws^2_W})lnq^2[
m^2_t(1+{1\over tan^2\beta})(\gamma^{\mu}P_L)\nonumber\\
&&+m^2_b(1+tan^2\beta)\{
(\gamma^{\mu}P_L)+2(\gamma^{\mu}P_R)\}]
\label{vertexgchi}
\eqa

\bqa
\Gamma^{Z}_{\mu}(\chi)&\to&
-({e\alpha\over48\pi M^2_Ws^3_Wc_W})lnq^2[
({3\over2}-s^2_W)
m^2_t(1+cot^2\beta)(\gamma^{\mu}P_L)
\nonumber\\
&&+m^2_b(1+tan^2\beta)\{({3\over2}-s^2_W)
(\gamma^{\mu}P_L)-2s^2_W(\gamma^{\mu}P_R)\}]
\label{vertexzchi}
\eqa

\bqa
\Gamma^{\gamma}_{\mu}(H)&\to&
-({e\alpha\over48\pi M^2_Ws^2_W})lnq^2
[m^2_t cot^2\beta(\gamma^{\mu}P_L)+
m^2_btan^2\beta\{(\gamma^{\mu}P_L)+2(\gamma^{\mu}P_R)\}
]
\label{vertexgh}
\eqa

\bqa
\Gamma^{Z}_{\mu}(H)&\to&
-({e\alpha\over48\pi M^2_Ws^3_Wc_W})lnq^2
[({3\over2}-s^2_W)m^2_t cot^2\beta (\gamma^{\mu}P_L)\nonumber \\
&&
+m^2_btan^2\beta)\{({3\over2}-s^2_W)(\gamma^{\mu}P_L) 
 -2s_W^2 (\gamma^{\mu}P_R)\}
]
\label{vertexzh}
\eqa
\noindent
where $P_{L,R}=(1\mp\gamma^5)/2$.

In the previous equations, we have retained not only the 
terms proportional to $m_t^2$ and to $m_b^2 tg^2\beta$,
as usually done (the latter ones become competitive for large 
$tg\beta$ values), but also those simply proportional
to $m_b^2$, that are usually discarded. 
Note that we did not retain SUSY masses 
inside the logarithm, being for the
moment only interested in the asymptotic energy limit. 
In principle, we could use a common reference mass M and
discard constant terms in the formulae. In fact, these possible 
constants will be thoroughly discussed in the final   
part of this paper. Thus, all the (bottom, top) mass terms contributing
the asymptotic logarithms has been retained and, as one sees, 
they are not vanishing and in principle numerically relevant, 
as one could easily verify by computing their separate effects 
on the various observables. This is, in principle, no surprise since
the corresponding terms in the SM were also, as we said, not negligible.
To be more precise, we write the "massive" SM vertices,
that were computed in Ref.\cite{mt}, simply
adding the terms proportional 
to $m_b^2$ that were neglected in that paper,
obtaining the expressions:

\bqa
\Gamma^{\gamma}_{\mu}(SM,~ massive)&\to&
-({e\alpha\over48\pi M^2_Ws^2_W})ln q^2
[m^2_t(\gamma^{\mu}P_L)+m^2_b(\gamma^{\mu}P_R)]\nonumber\\
&&-({e\alpha m^2_b\over48\pi M^2_Ws^2_W})ln q^2
[(\gamma^{\mu}P_L)+(\gamma^{\mu}P_R)]
\eqa

\bqa
\Gamma^{Z}_{\mu}(SM,~ massive)&\to&
-({e\alpha\over48\pi M^2_Ws^3_Wc_W})ln q^2
[({3\over2}-s^2_W)m^2_t(\gamma^{\mu}P_L)-s^2_Wm^2_b(\gamma^{\mu}P_R)]
\nonumber\\
&&-({e\alpha m^2_b\over48\pi M^2_Ws^3_Wc_W})ln q^2
[({3\over2}-s^2_W)(\gamma^{\mu}P_L)-s^2_W(\gamma^{\mu}P_R)]
\eqa

and adding Eqs.(\ref{vertexgchi})-(\ref{vertexzh}) we obtain
the total massive terms in the MSSM:

\bqa
\Gamma^{\gamma}_{\mu}(MSSM,~ massive)&\to&
-({e\alpha\over24\pi M^2_Ws^2_W})ln({q^2\over m^2_t})[
m^2_t(1+cot^2\beta)(\gamma^{\mu}P_L)\nonumber\\
&&+
m^2_b(1+tan^2\beta)\{(\gamma^{\mu}P_L)+2(\gamma^{\mu}P_R)\}]
\label{MSSMg}
\eqa

\bqa
\Gamma^{Z}_{\mu}(MSSM,~ massive)&\to&
-({e\alpha \over24\pi M^2_Ws^3_Wc_W})ln({q^2\over m^2_t})[
m^2_t({3\over2}-s^2_W)(1+cot^2\beta)(\gamma^{\mu}P_L)\nonumber\\
&&+
m^2_b(1+tan^2\beta)\{({3\over2}-s^2_W)(\gamma^{\mu}P_L)
-2s^2_W(\gamma^{\mu}P_R)\}]
\label{MSSMz}
\eqa

Notices that there exists a very simple practical
rule to move from the SM to the MSSM for what concerns
the asymptotic mass effects. One just multiplies the
$m^2_t$ term of the SM by $2(1+cot^2\beta)$ and the
$m^2_b$ one by $2(1+tan^2\beta)$
\footnote{We have checked that the signs of our vertices
agree with those of ref.\cite{Boulware} satisfying their
positivity prescription for the imaginary part
of the external fermion self-energies.}. This will
have practical consequences that will be fully illustrated in
the following Section III.

\section {Asymptotic expressions of the observables.}

After this preliminary discussion, we are now ready to compute 
the dominant asymptotic logarithmic terms in the various observables. 
For the massless SUSY partner sector of the MSSM, they
will only be produced by self-energies (the RG component) 
and by the vertices with $\chi^{\pm}, \chi^0$ shown   
in Fig.1, computed for massless
fermions (the "Sudakov-type" terms). 
For the massive sector they will be produced both by ($\chi$) mass
effects of Fig.1 and by ($H$) mass effects of Fig.2 as discussed in the
preceding Section.
Using the standard couplings 
conventions \cite{Roziek} leads to expressions for the
photon and $Z$ vertices that can be easily "projected" on the 
four gauge-invariant functions. From
the equations given in the Appendix B of Ref.\cite{log}
one can then derive the effect on various
observables. To save space and time, we omit
these intermediate steps and give directly the
latter expressions in the following equations. 
We have considered here both the case of
unpolarized production of the five "light" quarks and leptons 
and that of polarized initial
electron beams. The latter case would lead to the observation 
of a number of longitudinal polarization
asymmetries, whose properties have been exhaustively discussed 
elsewhere \cite{pol}. We
have considered for final quarks the overall hadronic 
production (symbol $5$) and that of the
separate bottom (symbol $b$), that exhibits interesting features 
that will be discussed. The overall results
shown in the following equations also include the SM effects 
previously computed \cite{log}, \cite{mt}.\\

 The various
terms are grouped in the following order:  first the RG(SM) with the
mass scale $\mu$, followed
by the linear and quadratic Sudakov (SM, W) terms,
the linear and quadratic Sudakov (SM, Z) terms and finally, in the case
of hadronic observables, the linear Sudakov term arising from
the quadratic $m^2_t$ contribution; then, in bold face,
the SUSY contributions, first the RG (SUSY) term with the
mass scale $\mu$, then the linear 
Sudakov (SUSY) term (scaled by the common mass $M$),
the linear massless Sudakov (SUSY) term arising 
from the quadratic $m^2_t$ 
contribution (scaled by a common mass $M'$) and in curly 
brackets the same term to which the $m_b^2\tan^2\beta$ contribution
is added for $\tan\beta=40$. 
This was done
in order to show precisely the difference
between the total SM prediction and the total SUSY part.\\

\bqa
\sigma_{\mu}&=&\sigma^{B}_{\mu}
[1+{\alpha\over4\pi}
\{(7.72\,N-20.58
)\,ln{q^2\over\mu^2}+(35.27\, ln{q^2\over M^2_W}-4.59\, 
ln^2{q^2\over M^2_W})
\nonumber\\
&&
+(4.79\,  ln{q^2\over M^2_Z} -1.43\, ln^2{q^2\over M^2_Z})
\nonumber\\
&&
+{\bf (3.86\,N+7.75
)\,ln{q^2\over\mu^2}-10.02\, ln{q^2\over M^2}}\}]
\label{sigmu}\eqa

\bqa\label{ass}
A_{FB,\mu}&=&A^{B}_{FB,\mu}+{\alpha\over4\pi}\{(0.54\,N-5.90
)\,ln{q^2\over\mu^2}+(10.19\, ln{q^2\over M^2_W}-0.08\,
ln^2{q^2\over M^2_W})
\nonumber\\
&&
+(1.25\, ln{q^2\over M^2_Z} -0.004\,ln^2{q^2\over M^2_Z})
\nonumber\\
&&
+{\bf(0.27\,N+1.57
)\,ln{q^2\over\mu^2}-0.079\, ln{q^2\over M^2}}\}
\label{AFBmu}\eqa

\bqa
A_{LR,\mu}&=&A^{B}_{LR,\mu}+{\alpha\over4\pi}\{(1.82\,N-19.79
)\,ln{q^2\over\mu^2}+(30.76\, ln{q^2\over M^2_W}-3.52\, 
ln^2{q^2\over M^2_W})
\nonumber\\
&&
+(0.78\,ln{q^2\over M^2_Z} -0.17\,ln^2{q^2\over M^2_Z})
\nonumber\\
&&
+{\bf (0.91\,N+5.25
)\,ln{q^2\over\mu^2}-3.69\, ln{q^2\over M^2}}\}.
\label{ALRmu}\eqa

\bqa\label{sigm5}
\sigma_{5}&=&
\sigma^{B}_{5}[1+ { \alpha\over4\pi}
\{ (9.88\,N-42.66
)\, ln{q^2\over\mu^2}+(46.58\, ln{q^2\over M^2_W}-6.30\, 
ln^2{q^2\over M^2_W})
\nonumber\\
&&
+(7.25\,ln{q^2\over M^2_Z} -2.03\,ln^2{q^2\over M^2_Z})
-~1.21 \ln{q^2\over m^2_t}\nonumber\\
&&
+{\bf(4.94\,N+13.66
)\,ln{q^2\over\mu^2}-10.99\, ln{q^2\over M^2}-3.65 \{-5.21\}
 \ln{q^2\over M'^2}
}\}
\eqa

\bqa
A_{LR,5}&=&A^{0}_{LR,5}+{\alpha\over4\pi}\{(2.11N-22.95
)\ln{q^2\over\mu^2}\nonumber\\
&&+(24.07 \ln{q^2\over M^2_W}-3.12\ln^2{q^2\over M^2_W})
\nonumber\\
&&
+(1.63\ln{q^2\over M^2_Z}-0.55\ln^2{q^2\over M^2_Z})
-~0.53 \ln{q^2\over m^2_t}
\nonumber\\
&&
+{\bf (1.05\,N+6.09
)\,ln{q^2\over\mu^2}-3.63\, ln{q^2\over M^2}
-1.60 \{+0.44\}\ln{q^2\over M'^2}}\}\ ,
\label{ALR5t}\eqa

\bqa
\sigma_{b}&=&\sigma^{B}_{b}\{1+{\alpha\over4\pi}\{(10.88N-53.82
)\ln{q^2\over\mu^2}+(76.75 \ln{q^2\over M^2_W}-
7.10\ln^2{q^2\over M^2_W})
\nonumber\\
&&
+(11.98\ln{q^2\over M^2_Z}-2.45\ln^2{q^2\over M^2_Z})
-~8.42 \ln{q^2\over m^2_t}
\nonumber\\
&&
+{\bf (5.44\,N+16.61
)\,ln{q^2\over\mu^2}-11.82\, ln{q^2\over M^2}
-25.3\{-36.0\} \ln{q^2\over M'^2}}\}\ ,
\label{sigbt}\eqa

\bqa
A_{FB,b}&=&A^{B}_{FB,b}+{\alpha\over4\pi}\{(0.56N-6.13
)\ln{q^2\over\mu^2}\nonumber\\
&&+(17.23 \ln{q^2\over M^2_W}-0.31\ln^2{q^2\over M^2_W})
\nonumber\\
&&
+(0.96\ln{q^2\over M^2_Z}-0.08\ln^2{q^2\over M^2_Z})
-~0.36 \ln{q^2\over m^2_t}
\nonumber\\
&&
+{\bf (0.28\,N+1.63
)\,ln{q^2\over\mu^2}-0.38\, ln{q^2\over M^2}
-1.10\{+0.26\} \ln{q^2\over M'^2}}
\}
 \ .
\label{AFBbt}\eqa

\bqa
A_{LR,b}&=&A^{B}_{LR,b}+{\alpha\over4\pi}\{(1.88N-20.46
)\ln{q^2\over\mu^2}\nonumber\\
&&+(27.91 \ln{q^2\over M^2_W}-2.35\ln^2{q^2\over M^2_W})
\nonumber\\
&&
+(1.92\ln{q^2\over M^2_Z}-0.52\ln^2{q^2\over M^2_Z})
-~2.39 \ln{q^2\over m^2_t}
\nonumber\\
&&
+{\bf (0.94\,N+5.43
)\,ln{q^2\over\mu^2}-2.86\, ln{q^2\over M^2}
-7.16\{+2.57\} \ln{q^2\over M'^2}}
\},
\label{ALRbt}\eqa

\bqa
A_{b}&=&A^{0}_{b}+{\alpha\over4\pi}\{(1.41N-15.38
)\ln{q^2\over\mu^2}\nonumber\\
&&+(31.03 \ln{q^2\over M^2_W}-1.76\ln^2{q^2\over M^2_W})
\nonumber\\
&&
+(4.30\ln{q^2\over M^2_Z}-0.49\ln^2{q^2\over M^2_Z})
-~2.38 \ln{q^2\over m^2_t}
\nonumber\\
&&
+{\bf (0.71\,N+4.08
)\,ln{q^2\over\mu^2}-2.25\, ln{q^2\over M^2}
-7.14\{+3.18\} \ln{q^2\over M'^2}}\},
\label{AFBpbt}\eqa

In the previous equations $\sigma$ denotes cross sections, 
$A_{FB}$ forward 
backward asymmetries,
$A_{LR}$ longitudinal polarization asymmetries, 
$A_b$ the forward-backward 
polarization asymmetry \cite{afbpol}.
The various "subtracted" Born terms are defined in
Refs.\cite{log},\cite{mt}.\par
Eqs.(\ref{sigmu})-(\ref{AFBpbt}) are the main result of this paper. 
To better appreciate their 
message, we have
plotted in the following Figs.(3-11) the asymptotic terms, 
with the following convention:
for cross sections, we show the relative effect; for asymmetries, 
the absolute effect. To fix a
scale, we also write in the Figure captions 
the value of the (asymptotic) 
"Born " terms. The plots have been
drawn in an energy region between one and ten TeV. Higher values seem 
to us not realistic at the moment.
For lower values we feel that the asymptotic approximation might be 
"premature" for SUSY masses of a few
hundred GeV that we assumed, and we shall return to this point in 
the final discussion.\\

As one sees from Figs.(3-11), a number of clean conclusions can be 
drawn in the considered energy range. In particular:\\

1) The shift between the SM and the MSSM effects is systematically 
large and visible in all the considered observables 
at the  reasonably expected luminosity values (a few hundreds
of $fb^{-1}$ per year at LC or CLIC leading to an accuracy
close to the percent level). In all the  cross sections, 
this shift is dramatic, sometimes changing the
sign of the effect and increasing or decreasing its absolute
value by factors two-three. Similar
conclusions are valid for the set of polarized asymmetries;
for unpolarized asymmetries, the effect
is less spectacular, but  still visible. This decrease of
spectacularity has a  simple technical
reason: for unpolarized asymmetries, the SM squared logarithms 
are practically vanishing so that
only linear logarithms survive. The delicate cancellation 
mechanism between linear and quadratic
logarithms, that was deeply upset in the case of the other 
variables by the extra linear SUSY
logarithms, is  therefore absent in the unpolarized asymmetries case.\\

2) The pure RG logarithmic approximation, shown in Figs.(3-11),
is in general rather different from
the overall (RG + "Sudakov") one in a way that can be  energy
dependent. For all the considered observables 
with the exception of $A_{LR,\mu}$ and $
A_{LR,b}$ this difference remains large and 
measurable at the expected luminosity
in the "CLIC special" energy region (3-5 TeV). Therefore, 
approximating the
asymptotic logarithmic terms with the pure RG components for 
the considered processes would be a          
catastrophic theoretical error in the MSSM case, exactly like 
it would have been in the SM situation.\\

3) Looking at  the size of the effect, one notices that this 
must be separately discussed for each
specific observable at different energies. 
If one sticks to the CLIC energy 
region, one notices that for $\sigma_{\mu}$
the MSSM effect is now comparable 
(but of opposite sign) to the SM case, 
reaching values of a few percent.
For $\sigma_5$ the effect is 
now reduced from beyond the SM ten percent 
to a value oscillating around 
the few percent level.
For bottom production, the effect is 
strongly dependent on $\tan\beta$ and reaches values of more than ten
percent for $\tan\beta=40$. 
For the asymmetries, as one can see from Figs.(5,9)
the effect is sometimes increased and sometimes
reduced and is always remaining of the few percent size.
It seems therefore that in some cases SUSY makes the SM 
one-loop effect less "dangerous", in other
cases it reverses the situation. For $\sigma_5$ the reduction 
of the effect in the CLIC region
would guarantee a reasonable validity of the perturbative 
expansion; for bottom production, the
conclusion depends on the value of $\tan^2\beta$. 
Note, though, that for higher 
energies these conclusions might
change, as shown by the shape of the various curves. As a 
general comment, our feeling is that
in the TeV regime, for the MSSM, the 
validity of a one-loop perturbative expansion
is apparently safer than in the SM case, with the remarkable exception
of $\sigma_b$ in the large $\tan\beta$ case.

One final point remains to be discussed. Up to now
we have only considered the dominant asymptotic SUSY
terms in the 1 TeV-10 TeV range. For the SM case, 
it was seen \cite{log} that these were 
able to reproduce
with good accuracy (at the few percent level) the 
complete effect, and that in order to give a
more complete parametrization it was sufficient 
to add to the logarithmic terms a constant one,
depending on the observable and which can be determined
e.g. by a standard best fit procedure.
This was possible because in the SM there were no other 
free parameters left. In the MSSM case,
the situation is at the moment more complicated, since all 
the parameters of the model are nowadays
unknown (this might be no more a problem in a few years...). 
To try to get at least a feeling of
what could happen, we have devoted the last Section IV to 
the discussion of the simplest example that
we can provide, that of the SUSY Higgses effect. 
 Our aim 
is only that of trying to derive, in this case,
an extra constant asymptotic contribution.
This will be shortly discussed in what follows.\\

\section {A simple asymptotic fit for a SUSY effect }

The logarithmic terms that we have computed are supposed to 
be the dominant SUSY ones at asymptotic
energies. For realistic smaller energy regions, there might be other 
SUSY contributions that cannot be neglected. 
The simplest example is that of constant terms, 
whose presence would lead to an expansion
for a general cross section or asymmetry
of the kind :
\bqa
\frac{\sigma^{Born+(1~loop~ SUSY)}-\sigma^{Born}}{\sigma^{Born}} = 
\frac{\alpha}{4\pi} (c_{1,\sigma} \log\frac{q^2}{M^2} + c_{0, \sigma} + 
\cdots ), \\
A^{Born+(1~loop~ SUSY)}-A^{Born} = 
\frac{\alpha}{4\pi} (c_{1, A} \log\frac{q^2}{M^2} + c_{0, A} + \cdots).
\eqa
\noindent
where "Born" now includes the SM value.
Here, $c_0$, $c_1$ are in principle functions
of all the free parameters (mixing angles and masses) 
of the virtual contributions under consideration. The choice of the 
mass scale $M$ affects the definition of $c_0$ 
and will be discussed below.
The label ``$(1~loop~SUSY)$''
stands for a definite subset of one loop diagrams (e.g. SUSY Higgses 
exchange, SUSY gauginos exchange).

In the SM case, an analogous 
simple possibility was considered \cite{log},
\cite{mt} and it was shown 
that the resulting expression
was fitting the accurate results to quite a good (few permille) 
accuracy  also in an energy range between
500 GeV and one TeV, where in principle it might have been a "poor" 
approximation. This was interpreted as
a consequence of a "precocious" asymptotism in the SM case, 
where all the relevant masses are well below the TeV value.\\
In the MSSM, the situation might be worse if the SUSY masses 
are relatively heavy. Still, the possibility
of a simple parametrization, e.g. valid in the CLIC region, appears 
qualitatively motivated. The
practical investigation of this idea would require, 
in principle, a lengthy calculation given
the number of parameters of the models (masses, mixings...). 
The latter ones typically disappear in 
the asymptotic terms as obvious, but would reappear in subleading terms
like the constant $c_0$, as one can
easily check by calculation e.g. of the massless vertices.\\
In this short final Section, we have analyzed the simplest 
case of the SUSY Higgses contribution, whose asymptotic
expression we have derived.  What we want to
do is to isolate this effect and try to estimate its 
subleading constant term.\\
 
With this purpose, we
have considered all those hadronic observables to which 
the SUSY Higgses diagrams do contribute; the exact (not asymptotic)
expression of the observables at the one loop level is of the
kind:
\bqa
\frac{\sigma^{Born+SUSY Higgs}-\sigma^{Born}}{\sigma^{Born}} =  
\frac{\alpha}{4\pi} F_\sigma(q^2, \tan\beta, M_A) \\
A^{Born+SUSY Higgs}-A^{Born} = 
\frac{\alpha}{4\pi} F_A(q^2, \tan\beta, M_A) 
\eqa
where $\beta$ is the mixing angle related to the 
two Higgs vacuum expectation values, $M_A$ is the mass of the CP odd 
SUSY Higgs boson $A^0$ and the masses of the other SUSY Higgs particles
have been determined by means of the code FEYNHIGGS~\cite{feynhiggs}.

Away from resonances, the function $F_{\cal O}$ (${\cal O}=\sigma$ or $A$)
is expected to 
be 
\bq
F_{\cal O} \simeq c_{1, \cal O}(\tan\beta, M_A) \log\frac{q^2}{M_A^2}+
c_{0, \cal O}(\tan\beta, M_A)
\eq
We carefully analyzed the behaviour of the hadronic 
observables $\sigma_b$, $\sigma_5$, $A_{FB, b}$, $A_{LR, b}$, $A_{LR, 5}$ and 
$A_{b}$. As a representative example, we consider here in some details
the case of $\sigma_5$. \\
In Fig.(\ref{figc0c1good}), we plot the coefficients $c_0$ and $c_1$ 
as functions of $M_A$ at $\tan\beta=2.0$. We obtained them by fitting 
with a standard $\chi^2$ procedure the full computation of the 
diagrams in the energy range between $2$ and $10$ TeV.
As one can see, the maximum absolute error in the fit $\varepsilon$
defined as 
\bq
\varepsilon(\tan\beta, M_A) = 
\max_{q^2} 
\left|F_{\cal O}(q^2, \tan\beta, M_A) - c_{1, \cal O}(\tan\beta, M_A) \log\frac{q^2}{M_A^2}-c_{0, \cal O}(\tan\beta, M_A)\right|
\eq
is completely 
negligible. 
This holds true as far as the fitting range does not include 
resonances. We checked that the region  
$\sqrt{s}> 2\ \mbox{TeV}$ and $M_A < 500 \ \mbox{GeV}$ 
is safe and perfectly reproduced for all the considered observables.\\
We have also tried to determine the possible dependence of 
$c_0$, $c_1$ on the free parameter $\tan\beta$ at fixed $M_A$.
From a numerical thorough analysis and motivated by the dependence 
on $\tan\beta$ of the diagrams with charged SUSY Higgses exchange,
we checked that for $\tan\beta>1$ the following functional form: 
\bq
c_{i, \cal O}(\tan\beta, M_A) = 
c^+_{i, \cal O}(M_A) \tan^2\beta +  
c^-_{i, \cal O}(M_A) \cot^2\beta 
\eq
reproduces perfectly the exact calculation
with mildly $M_A$ dependent coefficients $c^\pm_{i, \cal O}$.
The plot of $c^\pm_i$ in the case of $\sigma_5$ are shown in
Fig.(\ref{figcpcm}) where we remark that 
the coefficients of the logarithm $c_1^\pm$ are, as expected, 
roughly independent on $M_A$.
The remarkable (in our
opinion) fact is that the analytic parametrization reproduces the exact 
numerical calculation practically identically, as seen in 
Fig.(13).\\
It should be added that a similar parametrization 
in the energy region from 500 GeV to 1 TeV would be
much less satisfactory, and much more $M_A$-dependent. 
Just to give an example, we show in Fig.~(\ref{figc0c1bad})
what happens in the case of $\sigma_5$ at $\tan\beta=2.0$. 
Due to a resonance at about $\sqrt{q^2}=2 m_t$ in the vertex with 
two top quark lines and a single charged Higgs, the simple 
logarithmic representation of the effect is not accurate
and, in particular, the fitted coefficient $c_1$ is far
from its asymptotic value.

The lesson that we learn from
this example is, therefore, that a priori one can 
expect to be able to reproduce with simple analytical
expressions dominated by logarithms the MSSM prediction 
for all the relevant observables of the process
of $e^+e^-$ annihilation into fermion-antifermion in the TeV regime. 
This would be rather useful in the
(apparently probable) case of need of a perturbative expansion 
beyond the one-loop order, but could also
be used for the purposes of technical operations to be 
performed at one loop (QED ISR, for instance),
where the availability of such  a simple expression might 
be essential. In a forthcoming paper, we shall
develope a more complete study of this problem that also 
includes the other SUSY contributions of
"not SUSY-Higgses" type.\\

\section{Conclusions}

In this paper we have extended to the SUSY case the study
of the high energy behaviour of four-fermion processes
$e^+e^-\to f\bar f$, 
$f$ being a lepton or a light quark ($u,d,s,c,b$),
that we had previously performed in the SM case.
We have considered the asymptotic behaviour of the
four-fermion amplitudes at one loop and we have observed
that specific features differentiate the SUSY part from the SM part.\\

In both cases we first obtained the single logarithmic terms due
to photon and $Z$ self-energy contributions leading to the well-known
Renomalization Group effects.
However, in addition, we have found large logarithmic terms due to
non-universal diagrams, dubbed of "Sudakov-type". In SM there appear
linear logarithmic and quadratic logarithmic terms. In the SUSY
part there are only linear logarithmic terms. No quadratic logarithmic
terms are generated because of the specific spin structure
of the couplings to the SUSY partners appearing inside the diagrams.
In the Appendix we have given the explicit analytical asymptotic 
expressions of these various contributions (RG and Sudakov)
for both SM and MSSM.\\

The Sudakov terms arising in SUSY have additional
specific and very interesting features. Contrarily to SM where a
partial cancellation (at moderately high energies) appears between
linear and quadratic logarithmic terms, in the SUSY part linear 
terms are alone and remain important.
In particular they enhance the massive $m^2_t$, $m^2_b$
asymptotic contributions to $b\bar b$ production by factors
that depend on $tan^2\beta$ in a potentially visible way.\\

We have computed the effects of these asymptotic terms in the
various unpolarized and polarized observables, cross sections and
asymmetries. We have made illustrations for the high energy range
accessible to a future LC or CLIC, and we have shown the specific
behaviour of the SM and of the MSSM cases, emphasizing also the
large departure from what would have been expected taking only
the RG effects into account.\\

These results are important for the
tests of electroweak properties which will be performed at these
machines. They also indicate that for very high energies, if
a high accuracy is achievable, the one loop treatment
might be more reliable than in the SM case with the remarkable
exception of the $b\bar b$ cross section, 
for which a more complete two loop
calculation might be necessary, a situation which already occured
at the $Z$ peak ref.(\cite{BB}) \footnote{We are indebted
to R. Barbieri for a clarifying discussion
on this point}.
On another hand, for moderate energies (close to $1~TeV$), 
when SUSY masses fall in the few hundred GeV range so that
one is not yet in an asymptotic regime, 
we have shown that simple empirical formulae can 
reproduce the effect of subleading terms. We have made one illustration
with the SUSY Higgs effects on the total hadronic cross section. 
For a complete treatment much more work is required and this point 
is at present under investigation \cite{future}.\\

{\bf Acknowledgments}: This work has been partially
supported by the European
Community grant  HPRN-CT-2000-00149.

\newpage

\appendix

\section{Asymptotic logarithmic contributions in the MSSM}

\subsection{Universal ($\gamma,Z$-self-energy) SUSY contributions}

They arise from the bubbles (and associated tadpole diagrams) involving
internal L- and R- sleptons and squarks, charginos, neutralinos, 
as well as the charged and neutral Higgses and Goldstones
(subtracting the standard Higgs contribution):\\

\bqa
\tilde{\Delta}^{Univ}_{\alpha}(q^2)&\to&
{\alpha\over4\pi}(3+{16N\over9})(lnq^2)
\label{DAun}\eqa

\bqa
R^{Univ}(q^2)&\to&
-({\alpha\over4\pi s^2_W c^2_W})
[{13-26s^2_W+18s^4_W\over6}
+(3-6s^2_W+8s^4_W){2N\over9}](lnq^2)\nonumber\\
\label{Run}\eqa

\bqa
V^{Univ}_{\gamma Z}(q^2)=V^{Univ}_{Z\gamma}(q^2)&\to&
-({\alpha\over4\pi s_W c_W})
[{13-18s^2_W\over6}+(3-8s^2_W){2N\over9}](lnq^2)
\label{Vun}\eqa

\noindent
where N is the number of slepton and squark families. These terms
contribute to the RG effects.\\

\subsection{Non-universal SUSY contributions}

These are the contributions coming from triangle diagrams connected
either to the initial $e^+e^-$ or to the final $f\bar f$ lines, and
containing SUSY partners, sfermions $\tilde{f}$, charginos or
neutralinos $\chi_i$, or
SUSY Higgses (see Fig.1,2); external fermion self-energy diagrams
are added making the total contribution finite. These
non universal terms
consist in $m_f$-independent terms  and in $m_f$-dependent terms
(quadratic $m^2_t$ and $m^2_b$ terms).
In this subsection 
we write the $m_f$-independent terms appearing in each $e^+e^-\to
f\bar f$ process, the $m_f$-dependent terms being given, for
$e^+e^-\to b\bar b$, in the next subsection.\\

Contribution to $e^+e^-\to \mu^+\mu^-$

\bqa
\tilde{\Delta}_{\alpha,e\mu}(q^2)&\to&
({\alpha\over\pi}lnq^2)
({-5+6s^2_W\over 4c^2_W})
\label{DAnunmu}\eqa

\bqa
R_{e\mu}(q^2)&\to&({\alpha\over\pi}lnq^2)
({3-8s^2_W+12s^4_W\over 8s^2_Wc^2_W})
\label{Rnunmu}\eqa

\bqa
V_{\gamma Z, e\mu}(q^2)=V_{Z\gamma , e\mu}(q^2)
&\to&({\alpha\over\pi}lnq^2)
({9-30s^2_W+24s^4_W\over 16s_Wc^3_W})
\label{Vgznunmu}\eqa

Contribution to $e^+e^-\to d\bar d,~s\bar s,~ b\bar b$,

\bqa
\tilde{\Delta}_{\alpha,ed}(q^2)&\to&
({\alpha\over\pi}lnq^2)
({-7+8s^2_W\over 9c^2_W})
\label{DAnund}\eqa

\bqa
R_{ed}(q^2)&\to&({\alpha\over\pi}lnq^2)
({27-58s^2_W+64s^4_W\over 72s^2_Wc^2_W})
\label{Rnund}\eqa

\bqa
V_{\gamma Z, ed}(q^2)&\to&({\alpha\over\pi}lnq^2)
({45-146s^2_W+128s^4_W\over 144s_Wc^3_W})
\label{Vgznund}\eqa

\bqa
V_{Z\gamma , ed}(q^2)&\to&({\alpha\over\pi}lnq^2)
({81-210s^2_W+128s^4_W\over 144s_Wc^3_W})
\label{Vzgnund}\eqa

Contribution to $e^+e^-\to u\bar u, c\bar c$

\bqa
\tilde{\Delta}_{\alpha,eu}(q^2)&\to&
({\alpha\over\pi}lnq^2)
({-71+82s^2_W\over 72c^2_W})
\label{DAnunu}\eqa

\bqa
R_{eu}(q^2)&\to&({\alpha\over\pi}lnq^2)
({27-67s^2_W+82s^4_W\over 72s^2_Wc^2_W})
\label{Rgnunu}\eqa

\bqa
V_{\gamma Z, eu}(q^2)&\to&({\alpha\over\pi}lnq^2)
({63-200s^2_W+164s^4_W\over 144s_Wc^3_W})
\label{Vgznunu}\eqa

\bqa
V_{Z\gamma , eu}(q^2)&\to&({\alpha\over\pi}lnq^2)
({81-240s^2_W+164s^4_W\over 144s_Wc^3_W})
\label{Vzgnunu}\eqa\\

\subsection{Non-universal SUSY contributions, final $b\bar{b}$}

We now list the $m^2_t$ and $m^2_b$ dependent terms appearing in
$e^+e^-\to b\bar b$:

\bqa
\tilde{\Delta}_{\alpha,eb}(q^2)&\to&
\tilde{\Delta}_{\alpha,ed}(q^2)-
\frac{\alpha}{24\pi s_W^2} lnq^2
\left[
s_W^2\frac{m_t^2}{M_W^2}(1+2cot^2\beta)
+(3-s^3_W)(1+2tan^2\beta)
\frac{m_b^2}{M_W^2}
\right]
\eqa

\bqa
R_{eb}(q^2)&\to&
R_{ed}(q^2)+
\frac{\alpha}{16\pi s^2_W} ln q^2
\left[
(1-\frac{2s^2_W}{3}) \frac{m^2_t}{M^2_W}(1+2cot^2\beta)
 +(1+\frac{2s_W^2}{3})(1+2tan^2\beta) \frac{m_b^2}{M_W^2}
\right]
\eqa

\bqa
V_{\gamma Z, eb}(q^2)&\to&
V_{\gamma Z, ed}(q^2)
+{\alpha c_W\over24 \pi s_W} lnq^2
\left(
\frac{m^2_t}{M^2_W}(1+2cot^2\beta)-\frac{m_b^2}{M_W^2}(1+2tan^2\beta)
\right)
\eqa

\bqa
V_{Z\gamma , eb}(q^2)&\to&
V_{Z\gamma , ed}(q^2)
+{\alpha\over 16\pi s_W c_W} lnq^2
(1-\frac{2s^2_W}{3})
\left(
\frac{m^2_t}{M^2_W}(1+2cot^2\beta)-\frac{m_b^2}{M_W^2}(1+2tan^2\beta)
\right)
\eqa\\

\subsection{Universal SM contributions}

In order to allow an easy comparison of the above SUSY contributions
with the SM ones we now recall, in the next three subsections,
 the results obtained in \cite{log},
\cite{mt} for the same four gauge invariant functions.\\

\bq
\tilde{\Delta}^{(RG)}_{\alpha}(q^2,\theta)\to
{\alpha(\mu^2)\over12\pi}[{32\over3}N-21]ln({q^2\over \mu^2})
\label{daRG}\eq

\bq
R^{(RG)}(q^2,\theta)\to-{\alpha(\mu^2)\over4\pi
s^2_Wc^2_W}[({20-40c^2_W+32c^4_W\over9}N+{1-2c^2_W-42c^4_W\over6}]
ln({q^2\over \mu^2})
\label{RRG}\eq

\bq
V^{(RG)}_{\gamma Z}(q^2,\theta)
=V^{(RG)}_{Z\gamma}(q^2,\theta)
\to{\alpha(\mu^2)\over3\pi
s_Wc_W}[({10-16c^2_W\over6}N+{1+42c^2_W\over8}]
ln({q^2\over \mu^2})
\label{VRG}\eq

\subsection{Non-universal SM contributions, final fermions $f\neq b$}

\bqa
\tilde{\Delta}^{(S)}_{\alpha,lf}(q^2,\theta)&\to&
{\alpha\over4\pi}[6-\delta_u-2\delta_d]ln{q^2\over M^2_W}
+{\alpha\over12\pi}(\delta_u+2\delta_d)ln^2{q^2\over M^2_W}
+{\alpha(2-v^2_l-v^2_f)\over64\pi s^2_Wc^2_W}[3ln{q^2\over M^2_Z}
-ln^2{q^2\over M^2_Z}]\nonumber\\
&&-{\alpha\over2\pi}[(ln^2{q^2\over M^2_W}+2ln{q^2\over
M^2_W}ln{1-cos\theta\over2})(\delta_{\mu}+\delta_d)
+(ln^2{q^2\over M^2_W}+2ln{q^2\over
M^2_W}ln{1+cos\theta\over2}))\delta_u]\nonumber\\
&&-{\alpha\over256\pi Q_f
s^4_Wc^4_W}[(1-v^2_l)(1-v^2_f)(ln{q^2\over
M^2_Z}ln{1+cos\theta\over1+cos\theta})]\nonumber\\&&
\eqa

\bqa
R^{(S)}_{lf}(q^2,\theta)&\to&
-{3\alpha\over4\pi s^2_W}[2c^2_W-\delta_{\mu}-(1-{s^2_W\over3})\delta_u
-(1-{2s^2_W\over3})\delta_d]ln{q^2\over M^2_W}\nonumber\\
&&
-{\alpha\over4\pi s^2_W}
[\delta_{\mu}+(1-{s^2_W\over3})\delta_u+(1-{2s^2_W\over3})\delta_d]
ln^2{q^2\over M^2_W}\nonumber
\\
&& 
-{\alpha(2+3v^2_l+3v^2_f)\over64\pi s^2_Wc^2_W}[3ln{q^2\over M^2_Z}
-ln^2{q^2\over M^2_Z}]\nonumber\\
&&
+{\alpha c^2_W\over2\pi s^2_W}[(ln^2{q^2\over M^2_W}+2ln{q^2\over
M^2_W}ln{1-cos\theta\over2})(\delta_{\mu}+\delta_d)
+(ln^2{q^2\over M^2_W}+2ln{q^2\over
M^2_W}ln{1+cos\theta\over2}))\delta_u]\nonumber\\
&&
+I_{3f}{\alpha\over2\pi s^2_Wc^2_W}[v_lv_fln{q^2\over
M^2_Z}ln{1+cos\theta\over1+cos\theta}]\nonumber\\&&
\eqa

\bqa
V^{(S)}_{\gamma Z,lf}(q^2,\theta)&\to&
{\alpha \over8\pi c_Ws_W}([3-12c^2_W+2c^2_W(\delta_u+2\delta_d)]
ln{q^2\over M^2_W}-[1+{2\over3}c^2_W(\delta_u+2\delta_d)]
ln^2{q^2\over M^2_W})\nonumber\\
&&
-[{\alpha v_l(1-v^2_l)\over128\pi s^3_Wc^3_W}+
{\alpha|Q_f|v_f\over8\pi s_Wc_W}]
[3ln{q^2\over M^2_Z}-ln^2{q^2\over M^2_Z}]\nonumber\\
&&
+{\alpha c_W\over2\pi s_W}
[(ln^2{q^2\over M^2_W}+2ln{q^2\over
M^2_W}ln{1-cos\theta\over2})(\delta_{\mu}+\delta_d)
+(ln^2{q^2\over M^2_W}+2ln{q^2\over
M^2_W}ln{1+cos\theta\over2}))\delta_u]\nonumber\\
&&
+I_{3f}{\alpha\over16\pi s^3_Wc^3_W}[v_f(1-v^2_l)ln{q^2\over
M^2_Z}ln{1+cos\theta\over1+cos\theta}]\nonumber\\&&
\eqa

\bqa
V^{(S)}_{Z\gamma,lf}(q^2,\theta)&\to&
{\alpha \over8\pi cs}([3-12c^2_W-2s^2_W(\delta_u+2\delta_d)]
ln{q^2\over M^2_W}-[1-{2\over3}s^2_W(\delta_u+2\delta_d)]
ln^2{q^2\over M^2_W})\nonumber\\
&&
-[{\alpha v_f(1-v^2_f)\over128\pi |Q_f|s^3_Wc^3_W}
+{\alpha v_l\over8\pi s_Wc_W}]
[3ln{q^2\over M^2_Z}
-ln^2{q^2\over M^2_Z}]\nonumber\\
&&
+{\alpha c_W\over2\pi s_W}
[(ln^2{q^2\over M^2_W}+2ln{q^2\over
M^2_W}ln{1-cos\theta\over2})(\delta_{\mu}+\delta_d)
+(ln^2{q^2\over M^2_W}+2ln{q^2\over
M^2_W}ln{1+cos\theta\over2}))\delta_u]\nonumber\\
&&
+{\alpha\over32\pi Q_f s^3_Wc^3_W}[v_l(1-v^2_f)ln{q^2\over
M^2_Z}ln{1+cos\theta\over1+cos\theta}]\nonumber\\&&
\eqa

\noindent
where $\delta_{\mu,u,d}=1$ for $f=\mu,u,d$ and 0 otherwise and
$v_l=1-4 \,s^2_W,\;\;v_f=1-4 \,|Q_f|\,s^2_W $.\par
In each of the above equations, we have successively added the
contributions coming from triangles containing one or two $W$,
from triangles containing one $Z$, from $WW$ box and finally from $ZZ$
box.\\

\subsection{Non-universal SM contributions, final $b\bar b$.}

For $b\bar b$ production there are additional SM contributions 
proportional to $m^2_t$ and $m^2_b$ arising 
from triangles involving $G^{\pm,0}$ or $H_{SM}$ lines
and Yukawa couplings involving $m_t$ or $m_b$ (those $m_b$ terms which
only come from the kinematics and give contributions
vanishing like $m^2_b/q^2$ have been safely neglected).\\

\vspace{0.5cm}

\bqa
\tilde{\Delta}_{\alpha,lb}(q^2)&\to&
\tilde{\Delta}_{\alpha,ld}(q^2)-{\alpha\over24\pi s^2_W}
(ln{q^2\over M^2})
~[~s^2_W({m^2_t\over M^2_W})
+(3-s^2_W)({m^2_b\over M^2_W})~]
 \label{daasbh}\eqa
\bqa
R_{lb}(q^2)&\to&R_{ld}(q^2)+{\alpha\over16\pi s^2_W}(ln{q^2\over M^2})
~[(1-{2s^2_W\over3})({m^2_t\over M^2_W})
+(1+{2s^2_W\over3})({m^2_b\over M^2_W})~]
\label{Rasbh}
\eqa
\bqa
V_{\gamma Z,lb}(q^2)&\to&V_{\gamma Z,ld}(q^2)
+{\alpha c_W\over24\pi s_W}(ln{q^2\over M^2})
~[~({m^2_t\over M^2_W})
-({m^2_b\over M^2_W})~]
\label{Vgzasbh}
\eqa

\bqa
V_{Z\gamma,lb}(q^2)&\to&
V_{Z\gamma,ld}(q^2)+
{\alpha\over16\pi s_W c_W}(ln{q^2\over M^2})(1-{2s^2_W\over3})
[({m^2_t\over M^2_W})
-({m^2_b\over M^2_W})~]
\label{Vzgasbh}\eqa

\subsection{Non-universal massive MSSM contributions, final $b\bar b$}

\vspace{0.5cm}

Finally we find interesting to sum up all the massive $m^2_t$ and
$m^2_b$ terms appearing in the MSSM (SM and SUSY non-universal massive
contributions to $e^+e^-\to b\bar b$). We remark that the net effect as
compared to the SM result is
a factor $2(1+cot^2\beta)$ for the $m^2_t$ term and a factor 
$2(1+tan^2\beta)$ for the $m^2_b$ one:

\bqa
\tilde{\Delta}_{\alpha,eb}(q^2)&\to&
\tilde{\Delta}_{\alpha,ed}(q^2)-
\frac{\alpha}{12\pi s_W^2} lnq^2
\left[
s^2_W\frac{m_t^2}{M_W^2}(1+cot^2\beta)
+(3-s^2_W)(1+tan^2\beta)
\frac{m_b^2}{M_W^2}
\right]
\eqa

\bqa
R_{eb}(q^2)&\to&
R_{ed}(q^2)+
\frac{\alpha}{8\pi s^2_W} ln q^2
\left[
(1-\frac{2s^2_W}{3}) \frac{m^2_t}{M^2_W}(1+cot^2\beta)
 +(1+\frac{2s_W^2}{3})(1+tan^2\beta) \frac{m_b^2}{M_W^2}
\right]
\eqa

\bqa
V_{\gamma Z, eb}(q^2)&\to&
V_{\gamma Z, ed}(q^2)
+{\alpha c_W\over12 \pi s_W} lnq^2
\left(
\frac{m^2_t}{M^2_W}(1+cot^2\beta)-\frac{m_b^2}{M_W^2}(1+tan^2\beta)
\right)
\eqa

\bqa
V_{Z\gamma , eb}(q^2)&\to&
V_{Z\gamma , ed}(q^2)
+{\alpha\over 8\pi s_W c_W} lnq^2
(1-\frac{2s^2_W}{3})
\left(
\frac{m^2_t}{M^2_W}(1+cot^2\beta)-\frac{m_b^2}{M_W^2}(1+tan^2\beta)
\right)
\eqa\\

\begin{figure}[p]
\vspace*{1cm}
\[
\epsfig{file=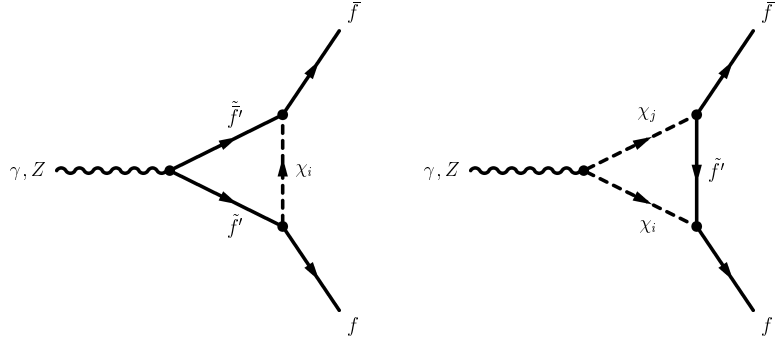,height=4.5cm}
\]
\vspace*{0.5cm}
\caption[1]{Triangle diagrams with SUSY partners exchanges
contributing to the asymptotic logarithmic behaviour in the energy;
$\chi_i$ represent either charginos or neutralinos.}
\label{tris}
\end{figure}\begin{figure}[p]

\[
\epsfig{file=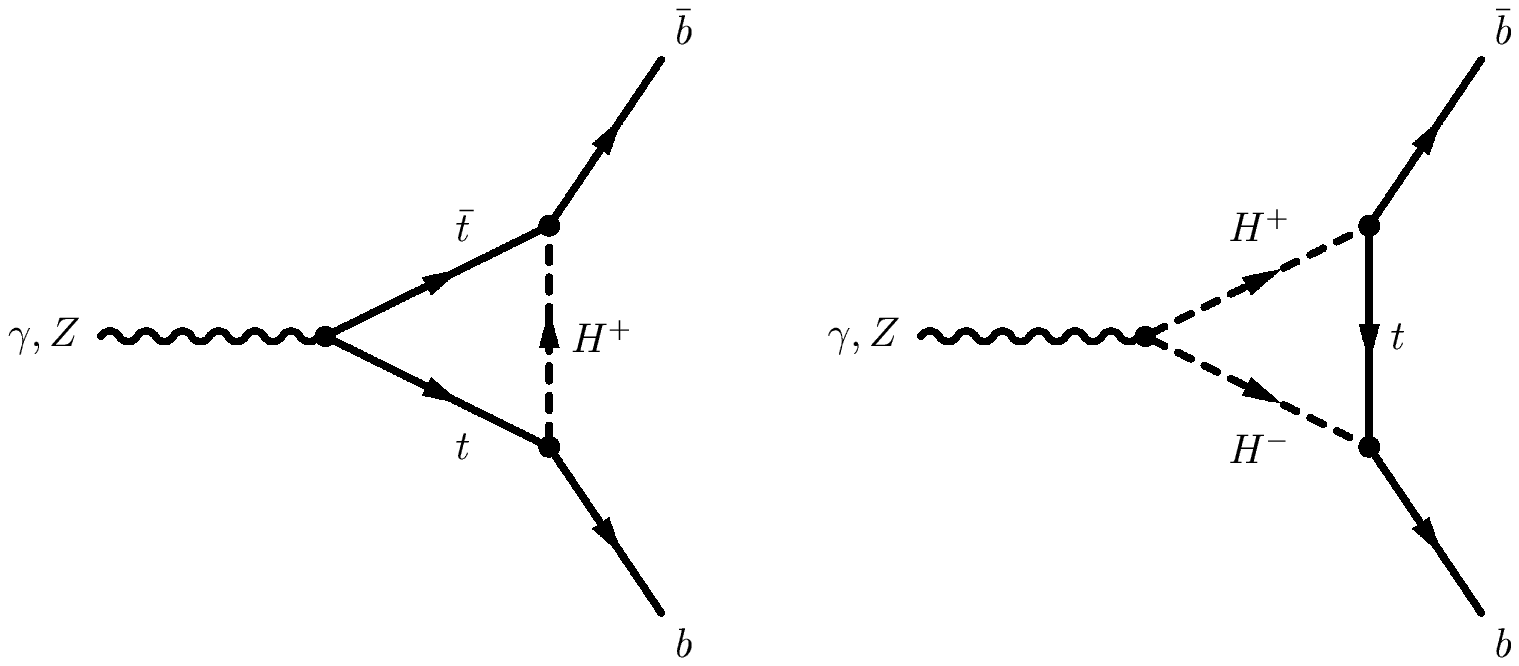,height=4.5cm}
\]
\vspace*{0.5cm}
\label{St}
\end{figure}\begin{figure}[p]
\vspace*{-2cm}
\[
\epsfig{file=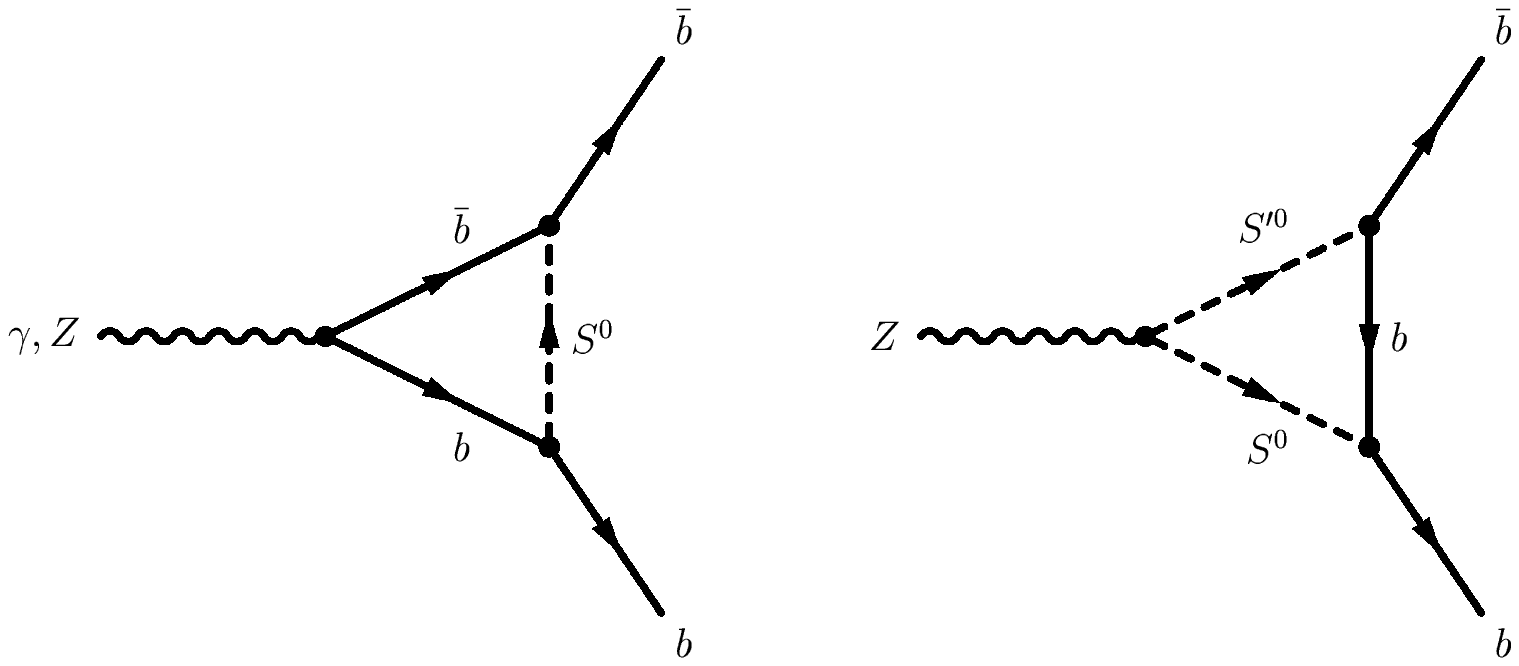,height=4.5cm}
\]
\caption[2]{Triangle diagrams of SUSY Higgs origin contributing 
to the asymptotic logarithmic behaviour in the energy; $S^0$
represent neutral Higgses $A^0$, $H^0$, $h^0$ or Goldstone $G^0$.}
\label{Sb}
\end{figure}
\newpage
\begin{figure}[p]
\vspace*{-2cm}
\[
\epsfig{file=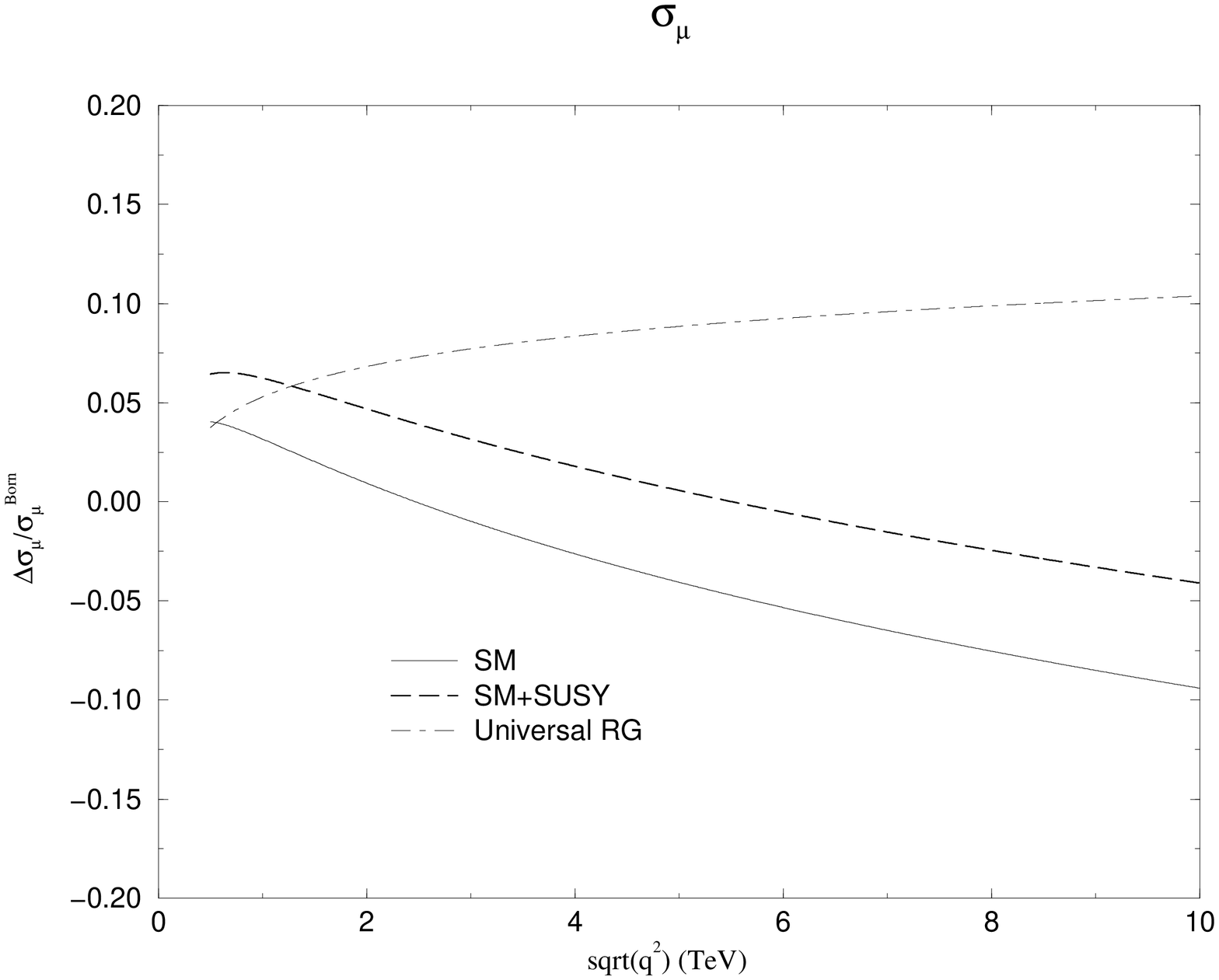,height=16cm,angle=-90}
\]
\caption[3]{Relative effects in $\sigma_\mu$ due to the asymptotic 
logarithmic terms. The Born expression for large $q^2$ is 
$111\ \mbox{pb}/(q^2/{\rm TeV}^2)$.
}
\label{sigmamu}
\end{figure}
\newpage
\begin{figure}[p]
\vspace*{-2cm}
\[
\epsfig{file=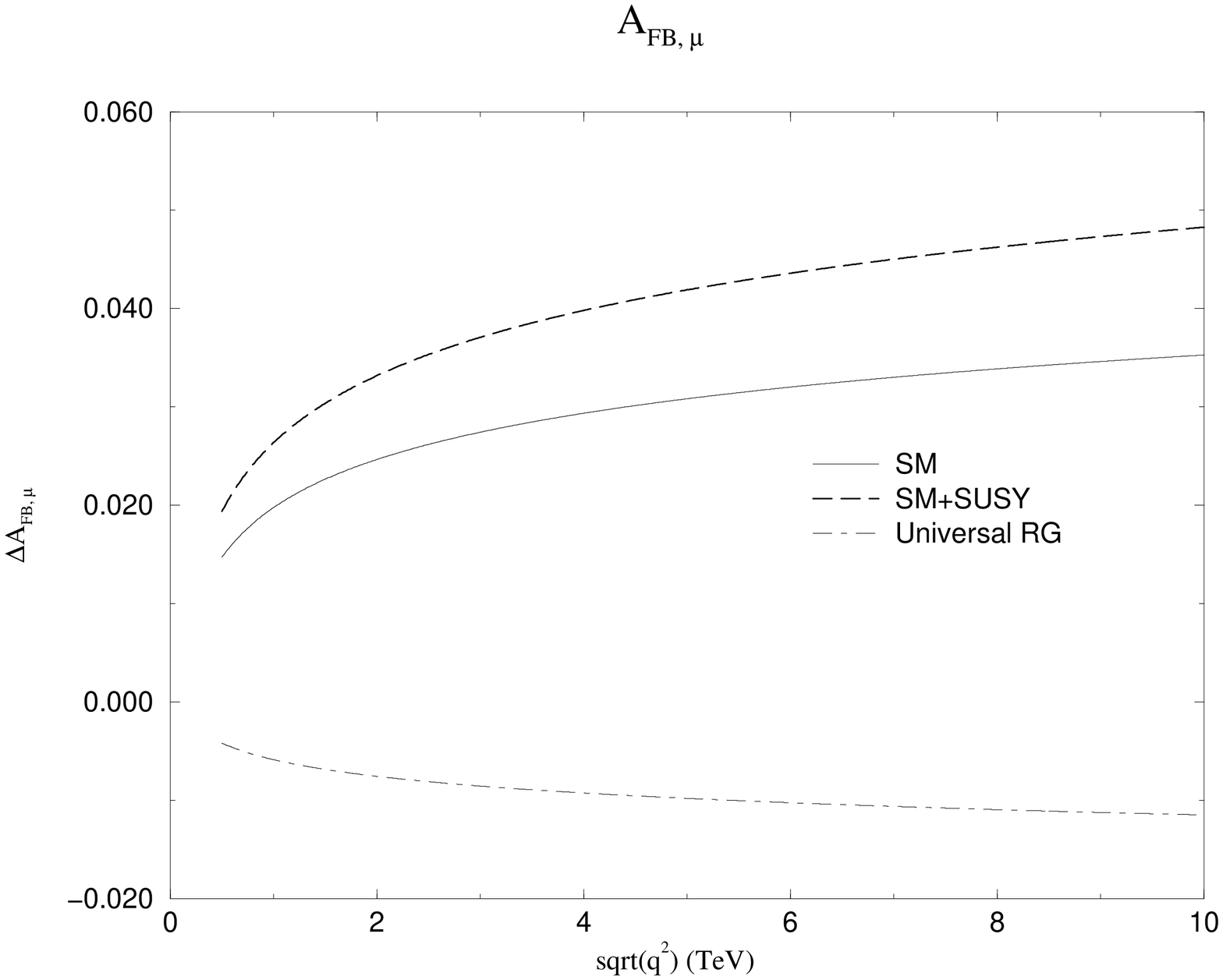,height=16cm,angle=-90}
\]
\caption[4]{Absolute effects in $A_{FB, \mu}$ due to the asymptotic 
logarithmic terms. The Born value for large $q^2$ is $0.47$.}
\label{afbmu}
\end{figure}
\newpage
\begin{figure}[p]
\vspace*{-2cm}
\[
\epsfig{file=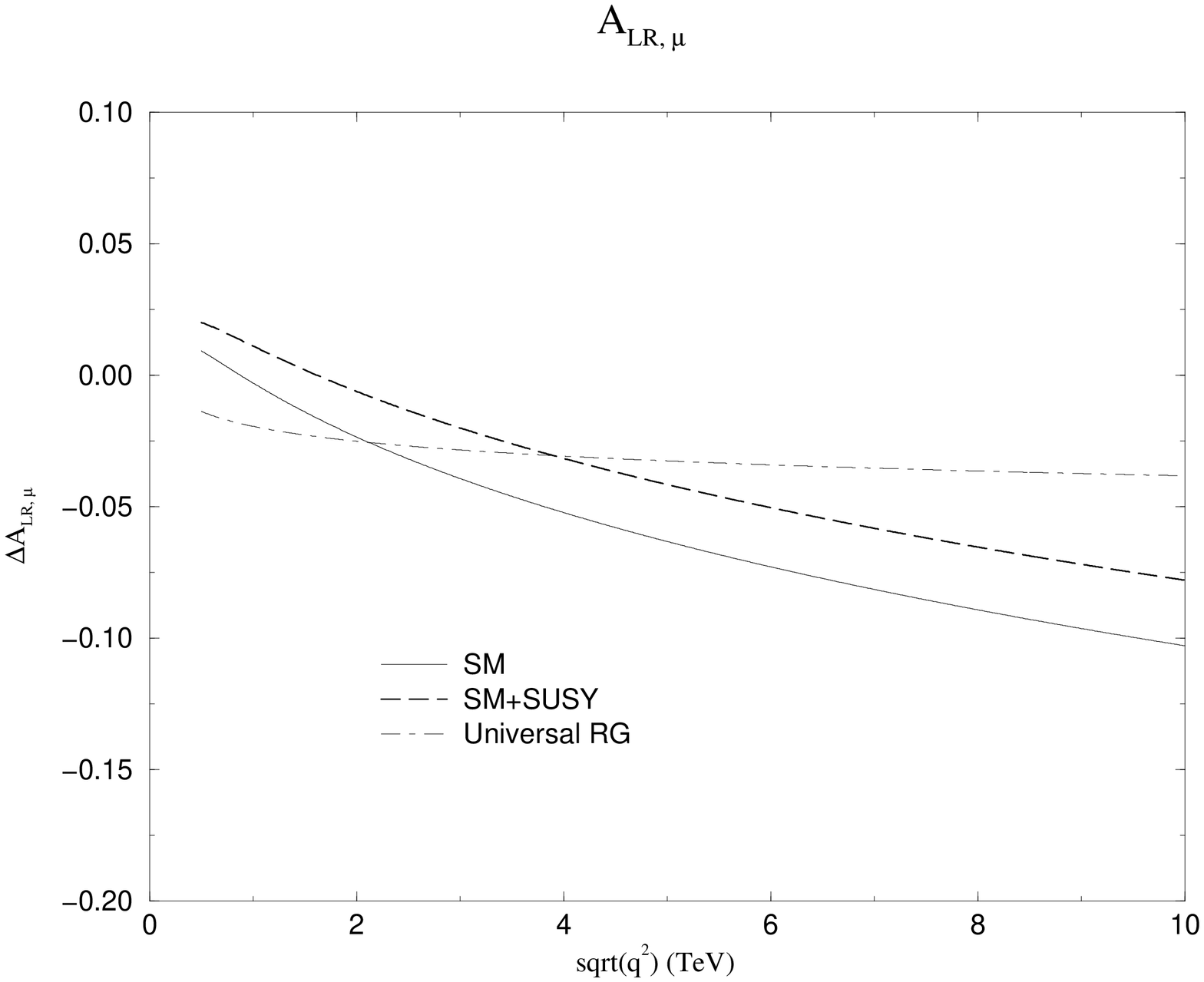,height=16cm,angle=-90}
\]
\caption[5]{Absolute effects in $A_{LR, \mu}$ due to the asymptotic 
logarithmic terms. The Born value for large $q^2$ is $0.063$.}
\label{alrmu}
\end{figure}
\newpage
\begin{figure}[p]
\vspace*{-2cm}
\[
\epsfig{file=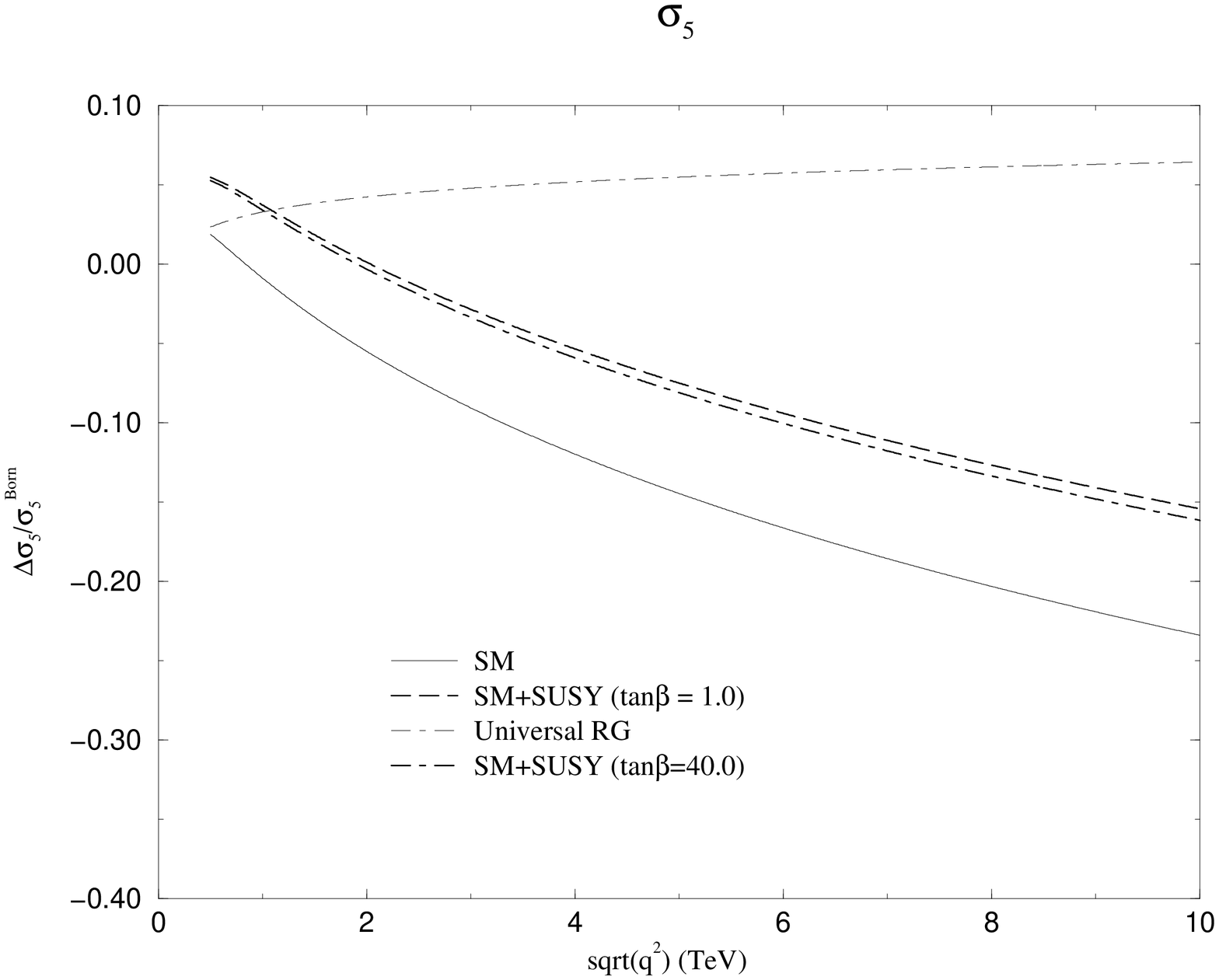,height=16cm,angle=-90}
\]
\caption[6]{Relative effects in $\sigma_5$ due to the asymptotic 
logarithmic terms.
The Born expression for large $q^2$ is 
$641\ \mbox{pb}/(q^2/{\rm TeV}^2)$.
}
\label{sigma5}
\end{figure}
\newpage
\begin{figure}[p]
\vspace*{-2cm}
\[
\epsfig{file=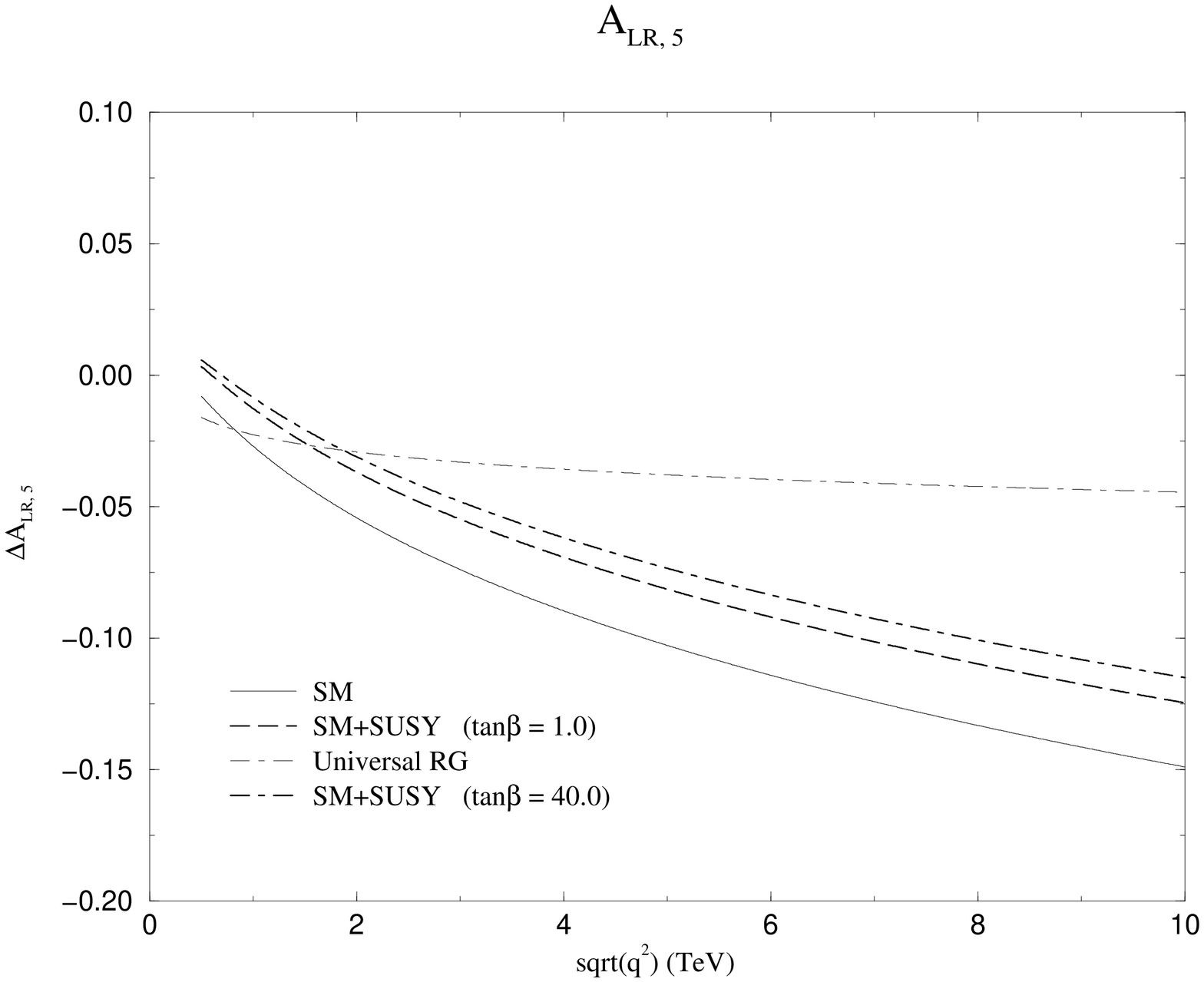,height=16cm,angle=-90}
\]
\caption[7]{Absolute effects in $A_{LR, 5}$ due to the asymptotic 
logarithmic terms. The Born value for large $q^2$ is $0.46$.}
\label{alr5}
\end{figure}
\newpage
\begin{figure}[p]
\vspace*{-2cm}
\[
\epsfig{file=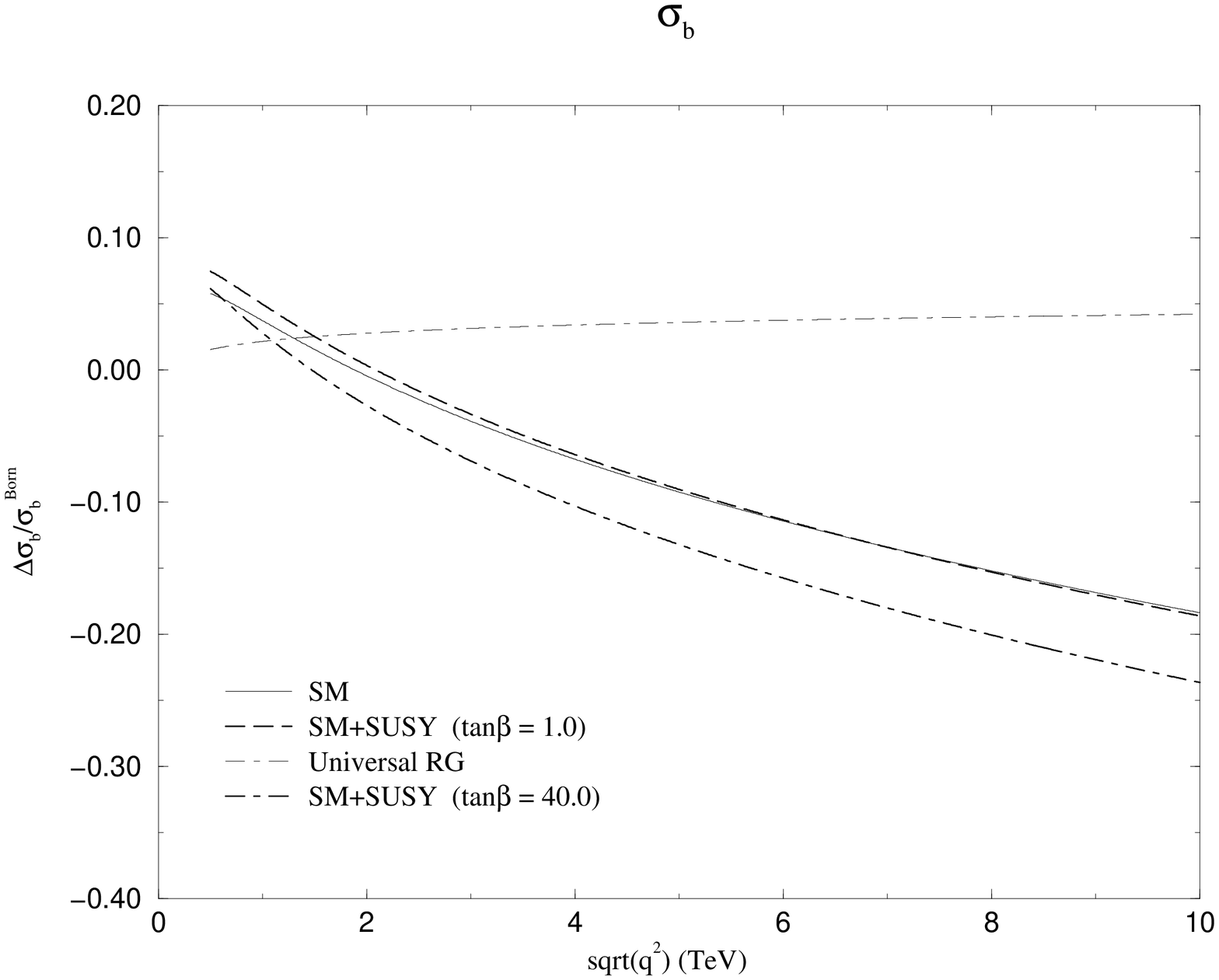,height=16cm,angle=-90}
\]
\caption[8]{Relative effects in $\sigma_b$ due to the asymptotic 
logarithmic terms.
The Born expression for large $q^2$ is 
$92\ \mbox{pb}/(q^2/{\rm TeV}^2)$.
}
\label{sigmab}
\end{figure}
\newpage
\begin{figure}[p]
\vspace*{-2cm}
\[
\epsfig{file=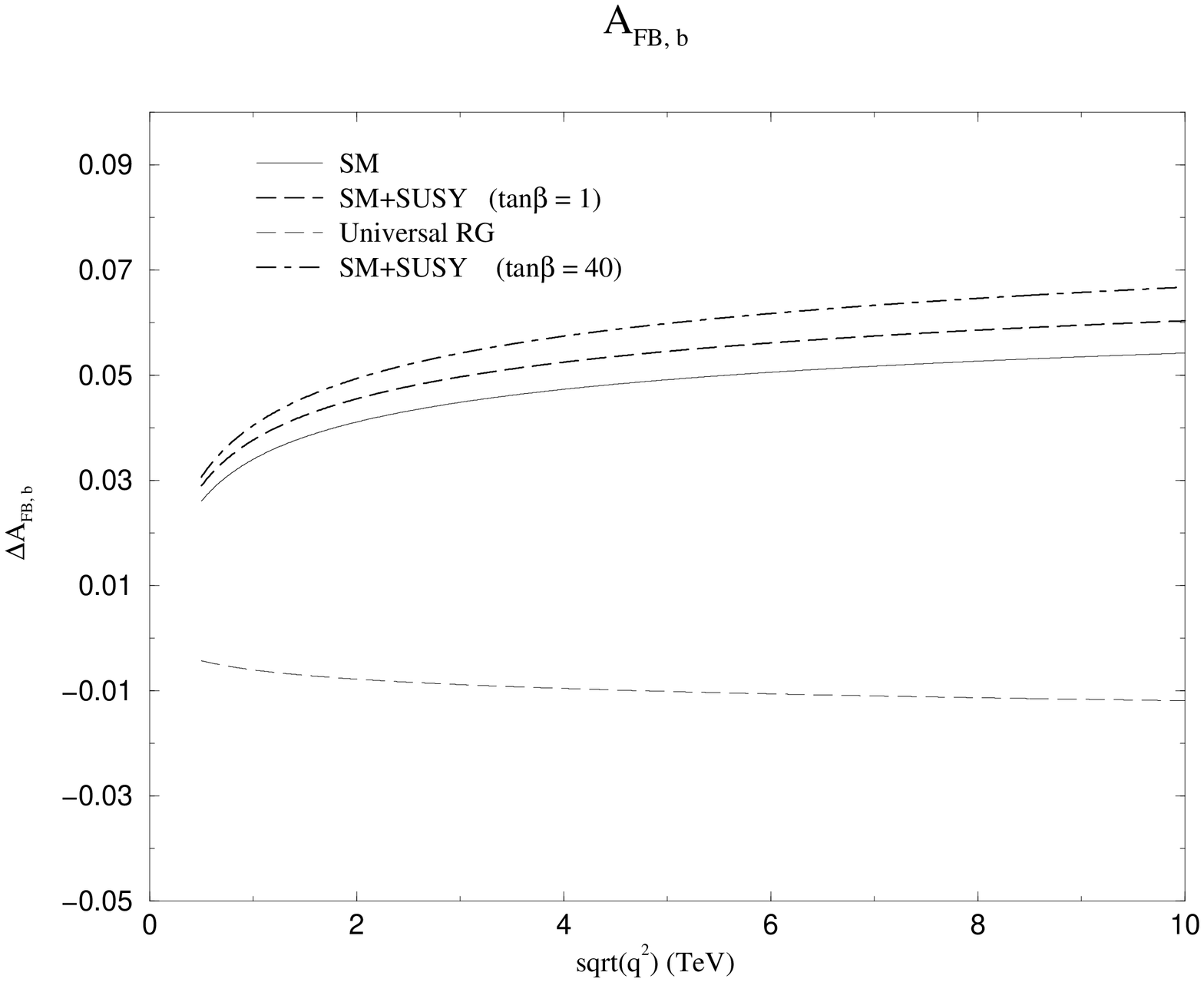,height=16cm,angle=-90}
\]
\caption[9]{Absolute effects in $A_{FB, b}$ due to the asymptotic 
logarithmic terms. The Born value for large $q^2$ is $0.64$.}
\label{afbb}
\end{figure}
\begin{figure}[p]
\vspace*{-2cm}
\[
\epsfig{file=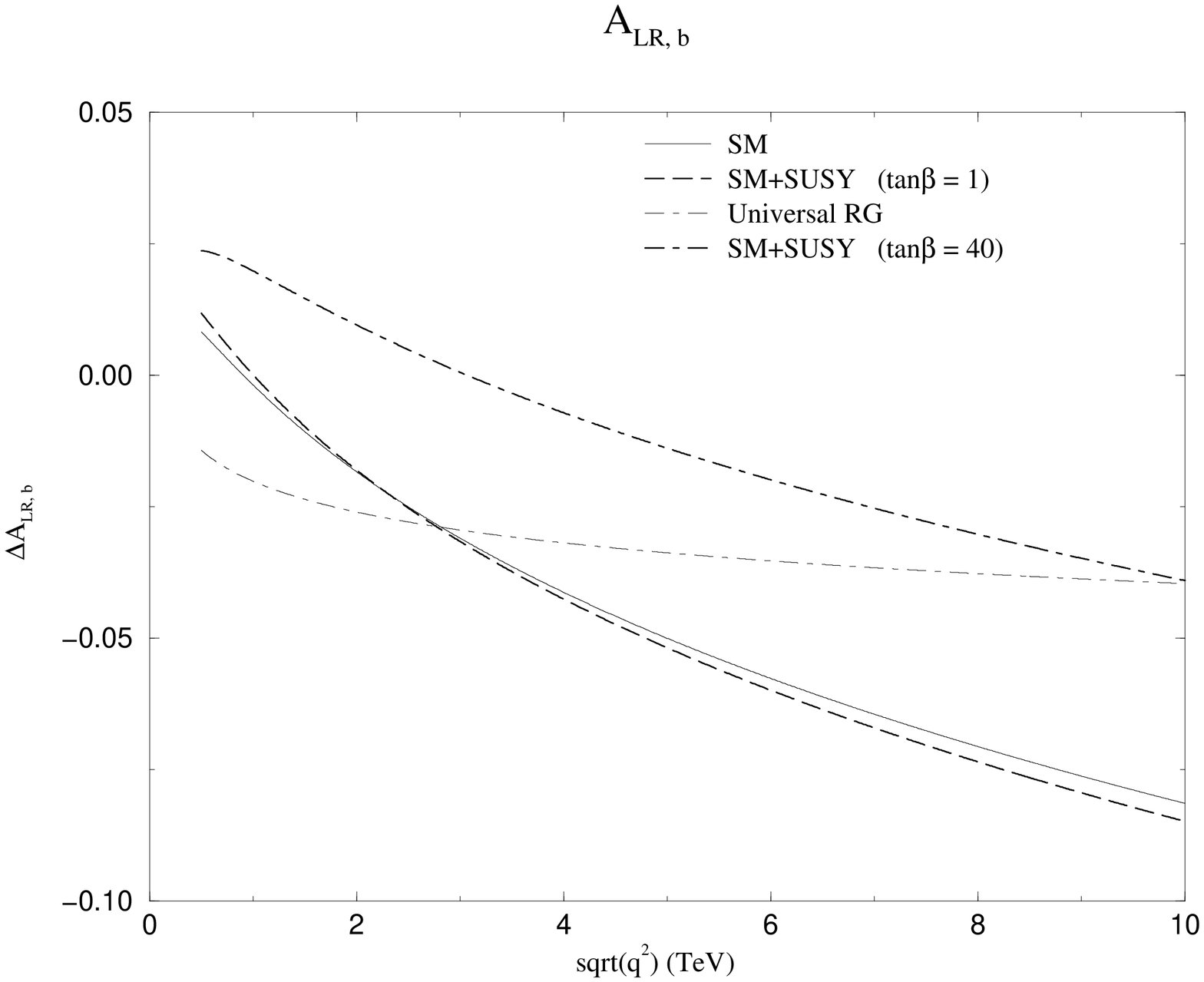,height=16cm,angle=-90}
\]
\caption[10]{Absolute effects in $A_{LR, b}$ due to the asymptotic 
logarithmic terms. The Born value for large $q^2$ is $0.62$.}
\label{alrb}
\end{figure}
\newpage
\begin{figure}[p]
\vspace*{-2cm}
\[
\epsfig{file=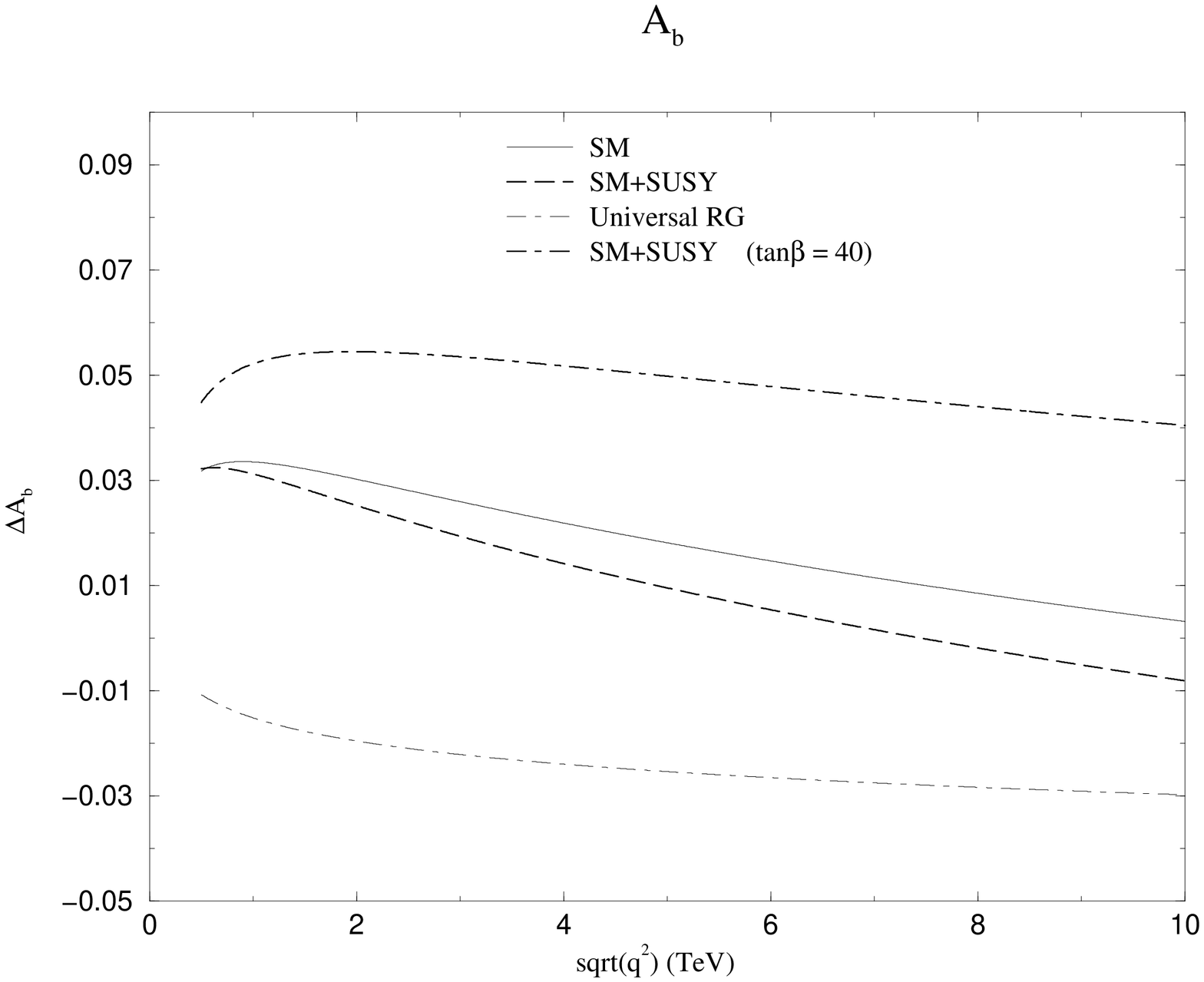,height=16cm,angle=-90}
\]
\caption[11]{Absolute effects in $A_{b}$ due to the asymptotic 
logarithmic terms. The Born value for large $q^2$ is $0.46$.}
\label{afbpolb}
\end{figure}
\newpage
\begin{figure}[p]
\vspace*{-2cm}
\[
\epsfig{file=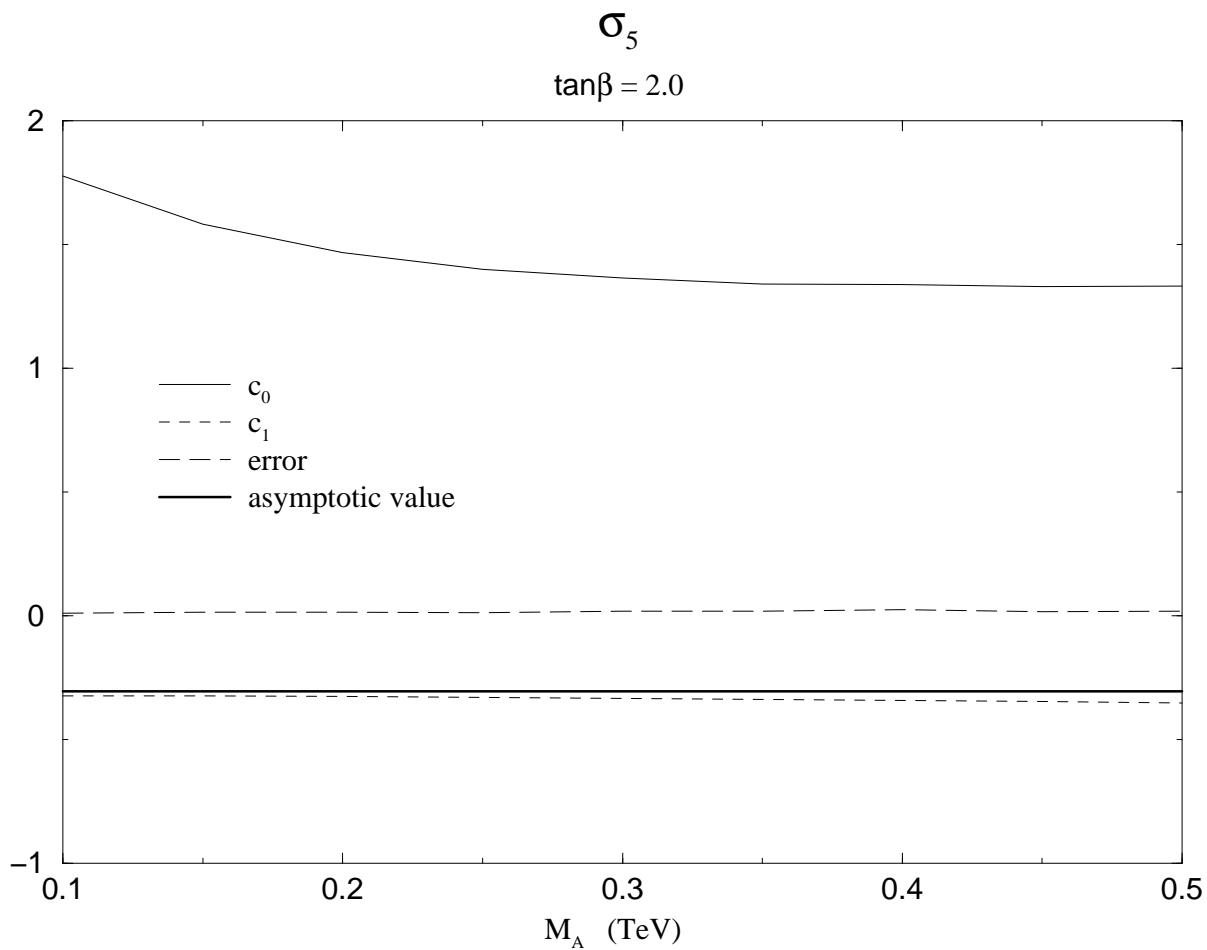,height=16cm,angle=-90}
\]
\caption[2]{
Effective parametrization of the SUSY Higgses effects in 
$\sigma_5$. The constants $c_0$ and $c_1$ are obtained by a $\chi^2$ fit
in the energy range between 2 and 10 TeV with $\tan\beta=2.0$.
The error quoted is the maximum absolute difference (with respect to 
$q^2$)
between the effective parametrization and the exact full calculation
and is always negligible. The constant $c_1$ is very near its analytical 
asymptotic value and as such is also roughly independent on $M_A$. 
On the other hand, the constant term $c_0$ is smoothly
dependent on $M_A$.
}
\label{figc0c1good}
\end{figure}
\newpage
\begin{figure}[p]
\vspace*{-2cm}
\[
\epsfig{file=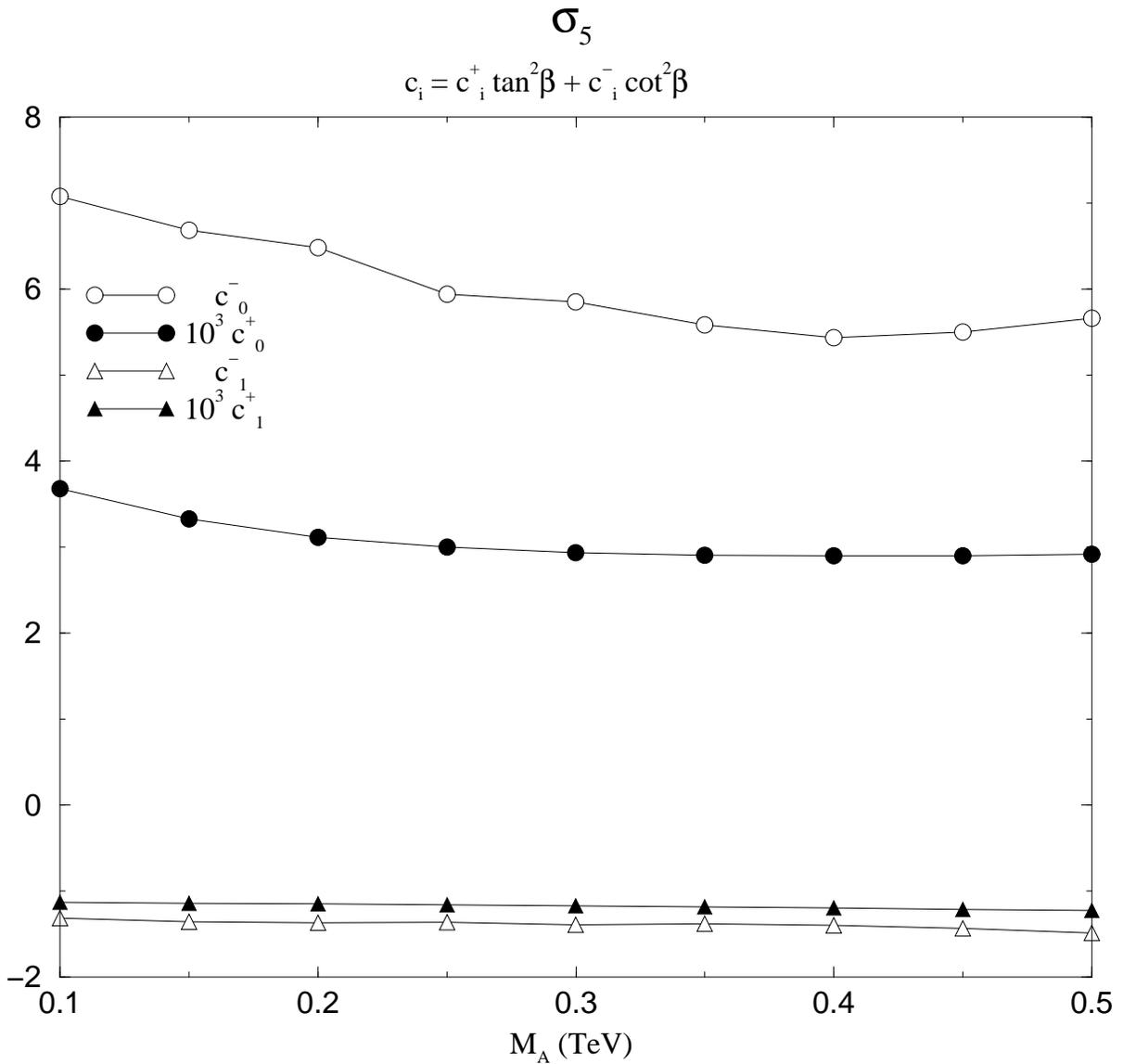,height=16cm,angle=-90}
\]
\caption[2]{
Dependence on $\tan\beta$ in the effective parametrization 
of the SUSY Higgses effects in $\sigma_5$. For each $M_A$, 
we determine the constants $c_{0,1}^\pm$ in 
$c_i = c_i^+ \tan^2\beta + c_i^-\cot^2\beta$. This functional form 
turns out to be perfectly matched by the exact calculation. We interpret
this fact as a dominance of the diagrams with exchange of charged SUSY
Higgses that have rigorously this dependence on $\tan\beta$.
Again, the coefficients of the logarithm, $c_1^\pm$ are 
 roughly independent on $M_A$. 
}
\label{figcpcm}
\end{figure}
\newpage
\begin{figure}[p]
\vspace*{-2cm}
\[
\epsfig{file=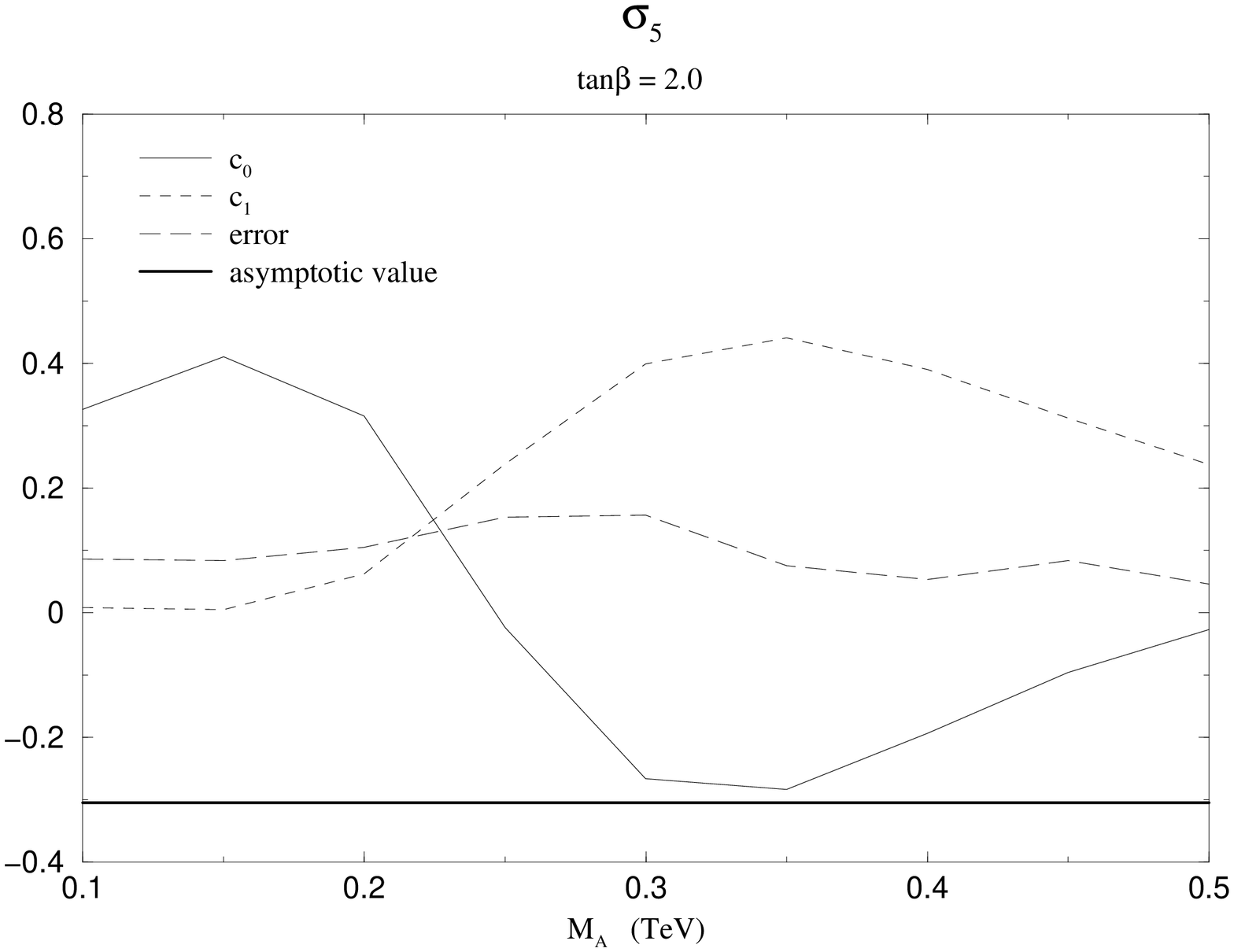,height=16cm,angle=-90}
\]
\caption[2]{
Effective parametrization of the SUSY Higgses effects in 
$\sigma_5$ in the energy range between 500 GeV and 1 TeV.
This region is definitely non asymptotic and 
the constants $c_0$, $c_1$ afforded by the best fit procedure
turn out to be strongly dependent on $M_A$ as discussed in the text.
}
\label{figc0c1bad}
\end{figure}
 

\begin{references}

\bibitem{log}  
M. Beccaria, P. Ciafaloni, D. Comelli, F.M. Renard and C. Verzegnassi,
Phys. Rev. {\bf D61},073005(2000).
%
\bibitem{mt}
M. Beccaria, P. Ciafaloni, D. Comelli, F.M. Renard and C. Verzegnassi, 
Phys.Rev. {\bf D61},011301(2000).

\bibitem{CC}
P. Ciafaloni, D. Comelli, Phys.Lett.{\bf B446},278(1999).

\bibitem{Sudakov} V. V. Sudakov, Sov. Phys. JETP 3, 65 (1956);
Landau-Lifshits:
Relativistic Quantum Field theory IV tome,  ed. MIR.

\bibitem{CCC} M. Ciafaloni, P. Ciafaloni, D. Comelli, 
Phys.Rev.Lett.{\bf 84},4810(2000); 
hep-ph/0004071;
hep-ph/0007096.

\bibitem{MSSM} H.P. Nilles, Phys.Rep. {\bf 110},1(1984); 
H.E. Haber and G.L.
Kane, Phys. Rep. {\bf 117},75(1985); 
R. Barbieri, Riv.Nuov.Cim. {\bf 11},1(1988);
R. Arnowitt, A, Chamseddine and P. Nath, "Applied N=1 Supergravity
(World Scientific, 1984); for a recent review see e.g. S.P. Martin,
hep-ph/9709356v3(1999).

\bibitem{nat} R. Barbieri, G.F. Giudice, Nucl.Phys.{\bf B306},63(1988).


\bibitem{LC} Opportunities
and Requirements for Experimentation at a Very High Energy
$e^{+}e^{-}$ Collider, SLAC-329(1928); Proc. Workshops on Japan
Linear Collider, KEK Reports, 90-2, 91-10 and 92-16;
P.M. Zerwas, DESY 93-112, Aug. 1993; Proc. of the Workshop on
$e^{+}e^{-}$ Collisions at 500 GeV: The Physics Potential, DESY
92-123A,B,(1992), C(1993), D(1994), E(1997) ed. P. Zerwas;
E. Accomando \etal\@ \prep{C299}{1998}{299}.

\bibitem{CLIC} " The CLIC study of a multi-TeV $e^+e^-$ linear
collider", CERN-PS-99-005-LP (1999).

\bibitem{Zsub} 
F.M.~Renard and C.~Verzegnassi, 
Phys. Rev. {\bf D52}, 1369 (1995), 
Phys. Rev. {\bf D53}, 1290 (1996);M. Beccaria, F. Renard, 
S. Spagnolo, C. Verzegnassi,
hep-ph/0002101; to appear in Phys.Rev.D.

\bibitem{DS} G. Degrassi and A. Sirlin, 
Nucl.~Phys. B {\bf 383}, 73 (1992); Phys. Rev. D {\bf 46}, 3104 (1992).


\bibitem{GUT} H. Georgi, H. Quinn and S. Weinberg, Phys.Rev.Lett.{\bf
32},451(1974); S. Dimopoulos and H. Georgi,
Nucl.Phys.{\bf B193},50(1981); S. Dimopoulos, S. Raby and F. Wilczek,
Phys. Rev.{\bf D24},1681 (1981); L. Iba\~ nez and GG. Ross,
Phys.Lett.{\bf 105B},439(1981).

\bibitem{Boulware}
M Boulware and D. Finnell, Phys.Rev.{\bf D44},2054(1991).

\bibitem{Roziek} J. Rosiek, Phys.Rev.{\bf D41},3464(1990); 
erratum hep-ph/9511250.

\bibitem{pol} F.M. Renard and C. Verzegnassi,
Phys.Rev.{\bf D55},4370(1997).


\bibitem{afbpol} A.Blondel, B.W.Lynn, F.M.Renard and C.Verzegnassi,
Nucl.Phys.{\bf B304},438(1988).

\bibitem{feynhiggs}
S. Heinemeyer, W. Hollik, G. Weiglein, 
Comput.Phys.Commun.124,76(2000); hep-ph/9812320.

\bibitem{BB} 
R. Barbieri, M. Beccaria, P. Ciafaloni, G. Curci, A. Vicere,
Nucl.Phys. {\bf B409}, 105 (1993); 
Phys.Lett. {\bf B288}, 95  (1992), Erratum-ibid. {\bf B312}, 511 (1993).

\bibitem{future}
M. Beccaria, F.M. Renard and C. Verzegnassi,
in preparation.





\vspace{1cm}


\end{references}
\end{document}